\documentclass[british]{article}
\usepackage[utf8]{inputenc}
\usepackage{verbatim}
\usepackage{multirow}
\PassOptionsToPackage{normalem}{ulem}
\usepackage{ulem}
\makeatletter

\pdfoutput=1
\usepackage{jheppub}
\renewcommand\[{\begin{equation}}
\renewcommand\]{\end{equation}}

\makeatother

\usepackage[british]{babel}

\begin{document}

\title{An exploratory study of heavy domain wall fermions on the lattice}

\collaboration{RBC and UKQCD Collaborations}
\author[a]{Peter Boyle,}
\author[b]{Andreas J\"uttner,}
\author[b,c]{Marina Krsti\'c Marinkovi\'c,}
\author[b]{Francesco Sanfilippo,}
\author[b]{Matthew Spraggs,}
\author[b]{Justus Tobias Tsang}

\affiliation[a]{School of Physics and Astronomy, University of
  Edinburgh, EH9 3JZ, Edinburgh, United Kingdom}
\affiliation[b]{School of Physics and Astronomy, University of
  Southampton, SO17 1BJ Southampton, United Kingdom}
\affiliation[c]{Theoretical Physics Department, CERN, Geneva,
  Switzerland}

\emailAdd{paboyle@ph.ed.ac.uk}
\emailAdd{a.juettner@soton.ac.uk}
\emailAdd{marina.marinkovic@cern.ch}
\emailAdd{f.sanfilippo@soton.ac.uk}
\emailAdd{m.spraggs@soton.ac.uk}
\emailAdd{j.t.tsang@soton.ac.uk}

\abstract{We report on an exploratory study of domain wall fermions
  (DWF) as a lattice regularisation for heavy quarks.  Within the
  framework of quenched QCD with the tree-level improved Symanzik
  gauge action we identify the DWF parameters which minimise
  discretisation effects.
  We find the corresponding effective 4$d$ overlap operator to be
  exponentially local, independent of the quark mass.
  We determine a maximum bare heavy quark mass of
  $am_h\approx 0.4$, below which the approximate chiral symmetry and
  O(a)-improvement of DWF are sustained. This threshold appears to be
  largely independent of the lattice spacing. Based on these findings,
  we carried out a detailed scaling study for the heavy-strange meson
  dispersion relation and decay constant on four ensembles with
  lattice spacings in the range $2.0-5.7\,\mathrm{GeV}$. We observe very mild
  $a^2$ scaling towards the continuum limit. Our findings establish a
  sound basis for heavy DWF in dynamical simulations of lattice QCD
  with relevance to Standard Model phenomenology.}

\maketitle

\section{Introduction \label{sec:intro}}
With LHCb and BESIII generating data, and Belle-II soon to start
production, increasingly accurate Standard Model (SM) predictions for
heavy flavour physics are dearly needed to constrain or
hopefully identify new physics. These predictions typically involve
matrix elements of the operators of the Weak Effective Hamiltonian among
hadronic states. As a result they require a non-perturbative approach,
making lattice QCD simulations crucial.

This is why in the last few years several approaches to implement
heavy quarks in simulations of lattice QCD have been proposed. Some of
these are based on an effective description of the heavy degrees of
freedom, such as the Non Relativistic treatment of the heavy quark
(NRQCD)~\cite{Lepage:1987gg,Thacker:1990bm} or Heavy Quark Effective
Theory (HQET)~\cite{Eichten:1989zv,Heitger:2003nj}, or on a non
relativistic re-interpretation of relativistic
discretisation~\cite{ElKhadra:1996mp,Christ:2006us,Lin:2006ur,Aoki:2001ra}.  More
recently collaborations have started treating the charm and bottom
quarks in the same relativistic framework used to discretise the light
quarks, e.g.~\cite{Follana:2006rc,Blossier:2009hg}.

However, simulations of full lattice QCD where both the physical light quarks
($u$, $d$ and $s$) and the charm or heavier quarks are
represented by the same discretisation are still rather
scarce.  The main reason is certainly that the relevant energy scales,
associated with the pion and the heavy quark masses respectively,
are computationally costly to reconcile. This is particularly true in
a fully relativistic and dynamical setup with controlled finite volume
and discretisation errors.  Simulations in which all quarks are
discretised in the same way have a number of advantages, though. For
instance, continuum flavour symmetries at finite lattice spacing
simplify many calculations. Moreover, only such a setup seems suitable
for the study of GIM-cancellation, which is an important ingredient in a
number of phenomenological
applications~\cite{Sachrajda:2015rea}. 

This paper is the second~\cite{Cho:2015ffa} in a series towards a
lattice phenomenology program with domain wall fermions
(DWF)~\cite{Kaplan:1992bt,Shamir:1993zy}, in particular M\"obius DWF
(MDWF)~\cite{Brower:2004xi,Brower:2005qw,Brower:2012vk}, as the
discretisation for light as well as heavy quarks.  Compared to Twisted
Mass~\cite{Frezzotti:2000nk}, DWF offer the attractive properties of
conserving both chiral and parity symmetries at finite lattice
spacing. Compared to HISQ fermions~\cite{Follana:2006rc}, a single
quark can be simulated without the need of taking the root of the
determinant to eliminate the different tastes, thus providing a
theoretically clean regularisation.

Since we enter mostly uncharted territory with simulations of heavy
DWF (see also~\cite{Chen:2014hva,Cho:2015ffa,Lin:2006vc}), we dedicate this paper
to the investigation of its basic properties.  We are particularly
interested in studying the approach of heavy-light meson observables
to the continuum limit. Our simulations have been carried out within
quenched QCD. This is computationally much cheaper than dynamical QCD and therefore
allows us to access a much wider range of lattice spacings
($a^{-1}\approx 2.0-5.7$\,GeV).
While the quenched approximation is certainly not suited for
making phenomenologically relevant predictions, we expect it to share
a number of properties with the unquenched case. Most importantly, we
expect that the continuum limit scaling observed in the quenched
theory over a large range of lattice spacings will be qualitatively
the same as in the dynamical theory. Such information is particularly
valuable given that for phenomenologically relevant simulations only
dynamical ensembles at coarser lattice spacings are currently
available.

The rest of the paper is organised as follows: in
section~\ref{sec:Strategy} we outline the overall computational
strategy followed in this paper, report on the properties of the
generated quenched gauge field ensembles, define the quantities that
we compute and discuss several more technical aspects of our
computation.  In section~\ref{sec:Tuning-DWF-for} we describe the
tuning of the MDWF parameters.  This is followed in
section~\ref{sec:Continuum-limit} by a study of the continuum limit
scaling of the dispersion relation and decay constants. In
section~\ref{sec:Conclusion} we draw our conclusions.  In the appendix
we provide supplementary material, in particular the numerical values
for all data underlying the analysis.

\section{Computational strategy and setup}\label{sec:Strategy}

\subsection{Strategy}

The main purpose of this work is to gain a qualitative understanding
of discretisation effects of heavy MDWF. To this end, we study the
MDWF parameter space and the heavy quark mass dependence of basic
heavy-heavy and heavy-strange meson matrix elements and the energy as
the cutoff is varied. Simulations of the quenched theory allow us to
adopt algorithms (over-relaxed~\cite{Creutz:1987xi,Kennedy:1985nu} heat-bath~\cite{Cabibbo:1982zn}) that are, compared to the
algorithms used with dynamical quarks (Hybrid Monte Carlo~\cite{Duane:1987de}),
computationally much cheaper. To some extent the problem of critical
slowing down~\cite{DelDebbio:2004xh,Schaefer:2010hu,Flynn:2015uma} can
therefore be circumvented by {\it brute force}.  This enables us to
probe finer lattice spacings than those affordable in dynamical
simulations and check the scaling of the theory towards the continuum
limit in more detail.
In order to reduce simulation costs further, a relatively small
physical lattice volume of $L\approx 1.6$\,fm was considered. The
volume was kept approximately constant while decreasing the lattice
spacing. Since the finite size effects in physical quantities
are then constant across all simulated lattice spacings, cut-off
effects can be studied in detail.

An important point addressed in this study concerns the residual
chiral symmetry breaking of MDWF. The restoration of chiral symmetry
in the massless limit is crucial to the simulation of QCD on the
lattice, and is also responsible for automatic
$\mathcal{O}(a)$-improvement, which is especially important when
studying heavy quark physics. 
{
In our notation, the five dimensional MDWF action is
$ S^5 = \bar{\psi} D^5_{MDWF} \psi $,
where
\begin{eqnarray}\label{eq:MDWFop}
D^5_{MDWF}
&=&
\left(
\begin{array}{cccccc}
  \tilde{D} & - P_- & 0& \ldots & 0 &  am P_+ \\
-  P_+  & \ddots &  \ddots & 0      & \ldots &0 \\
0     & \ddots &  \ddots & \ddots & 0      &\vdots \\
\vdots& 0      &  \ddots & \ddots & \ddots & 0\\
0     & \ldots &    0    &  \ddots& \ddots & - P_- \\
am P_- & 0      & \ldots  &  0     &   - P_+   
& \tilde{D}
\end{array}
\right)\,,
\end{eqnarray}
and we define
\begin{equation}
D_+ = (b D_W + 1)\,,\;\;
D_-= (1-c D_W )\;\;{\rm and}\;\;
\tilde{D} = (D_-)^{-1} D_+\,,
\end{equation}
with the usual chiral projectors $P_\pm=\frac 12(1\pm\gamma_5)$ and the Wilson matrix 
$D_W(M) = M+4 - \frac{1}{2} D_{\rm hop},$ 
where
$D_{\rm hop} = (1-\gamma_\mu) U_\mu(x) \delta_{x+\mu,y} +
              (1+\gamma_\mu) U_\mu^\dagger(y) \delta_{x-\mu,y}$
 acting in 4$d$.}
Besides the bare quark mass $am$, MDWF have two further input
parameters that need to be specified in each simulation: the extent of
the fifth dimension $L_s$ and the {\it domain wall height} parameter
$M=-M_5$, respectively.  More
specifically, $M_5$ is the negative mass parameter in the
4-dimensional Wilson Dirac operator that resides in the 5-dimensional
MDWF Dirac operator. Since both $L_s$ and $M_5$ are parameters of the discretisation
rather than of QCD we have some freedom in varying them. 
In the limit $L_s\to\infty$ and
with the Wilson kernel this formalism coincides with the overlap
formulation~\cite{Neuberger:1997bg,Neuberger:1997fp} and allows for
the simulation of a four-dimensional chirally symmetric theory (in the
limit of massless quarks) that is free of doublers. When $L_{s}$ is
finite however, chiral symmetry remains broken by a small
amount.\footnote{In fact, one expects residual chiral symmetry
  breaking to decrease {$\propto e^{-\alpha s}$ with some real and
    positive $\alpha$ when the Wilson kernel  has no zero modes
    (c.f.~\cite{Antonio:2008zz})}} This can be quantified by measuring
the amount of additive quark mass renormalisation, also known as
\emph{residual mass} $m_{\mathrm{res}}$ (defined later
in~\ref{sec:Observables}). For a given extent of the fifth dimension,
the parameter $M_{5}$  sets the
scale for the exponential localisation of the chiral modes of the
fermionic fields at the boundaries of the 5th dimension. The decay rate of the physical mode away from the
boundary is however also modified by the presence of an explicit quark
mass term, and care must be taken in order to maintain the localisation of
the physical modes on the
boundary~\cite{Kaplan:1992bt,Jansen:1992tw,Kaplan:2009yg,Christ:2004gc,Lin:2006vc,Liu:2003kp}. As
we will see, this becomes particularly crucial for {\it heavy} input
quark masses. We will study how the choice of a heavy quark mass $am=am_h$ and $M_5$ changes
the ultra-violet properties of the discretisation. In the following we
chose an extent $L_s=12$ of the fifth dimension, which guarantees a
small value of $m_{\rm res}$ for light quarks~\cite{Blum:2014tka}.
The particular choice of MDWF is the same implementation as the one used in~\cite{Blum:2014tka}
with a M\"obius scale of $\alpha=b+c=2$.

\subsection{Ensemble generation}
We generated ensembles based on the tree-level Symanzik
improved~\cite{Curci1983a,Luscher:1984xn} gauge action with lattice
spacings in the range of 0.034--0.1\,fm.  The gauge configurations
have been produced with the heat-bath
algorithm~\cite{Cabibbo:1982zn,Creutz:1987xi,Kennedy:1985nu}. The
coarser three ensembles were generated using
CHROMA~\cite{Edwards:2004sx},\footnote{We added heathbath routines for the tree level Symanzik action
to CHROMA.} whereas for the finest lattice spacing
(which involved the highest computational cost) we recurred to a
faster implementation, especially optimised for IBM
BG/Q~\cite{Sanfilippo_code}. In tables~\ref{tab:runpars}, \ref{tab:lat_vol} and~\ref{tab:autocorr}
 we summarise the simulation parameters used and basic ensemble properties.

\begin{table}
  \centering{}%
  \begin{tabular}{|c||c|c|}
    \hline $\beta$ & $L/a$ & $N_{\mathrm{sweeps}}$\\ \hline 4.41 & 16 & 10~k\\ 4.66
    & 24 & 20~k\\ 4.89 & 32 & 600~k\\ 5.20 & 48 & 1.4~M\\ \hline
  \end{tabular}
  \caption{Coupling constant $\beta$, volume in lattice units
    ($N_{x}=N_{y}=N_{z}=L/a$, $N_{t}=2L/a$), and number of update sweeps
    ($N_{\mathrm{sweeps}}$) after thermalisation.  Each sweep consists of 1
    heat-bath combined with 8 over-relaxation steps.}
  \label{tab:runpars}
\end{table}

Lattice spacings have been determined at each simulated $\beta$ by
enforcing the Wilson-flow scale
$w_{0}$~\cite{Luscher:2010iy,Borsanyi:2012zs} to take its ``physical''
value, which we assumed to be
$w_{0}^{\mathrm{phys}}=0.17245(99)\,\mathrm{fm}$ as recently
determined in~\cite{Blum:2014tka}.\footnote{Note that this 
value differs from the one used in~\cite{Cho:2015ffa}, $w_0=0.176(2)$\,fm~\cite{Borsanyi:2012zs}.}
 We kept the physical volume fixed such that the spatial extent
remained at about $1.6\,{\rm
  fm}$ (cf. table~\ref{tab:lat_vol}).

\begin{table}
\centering{} %
\begin{tabular}{|c||cccc|}
\hline $\beta$ & Plaquette &$w_{0}/a$ & $a^{-1}[\mathrm{GeV}]$ &
$L[\mathrm{fm}]$\\\hline
4.41 & 0.62637(3)   & 1.767(3) & 2.037(08) & 1.550(6)\\
4.66 & 0.651421(12) & 2.499(8) & 2.861(09) & 1.655(5)\\
4.89 & 0.671257(5)  & 3.374(11) & 3.864(12) & 1.634(5)\\
5.20 & 0.694149(4)  & 5.007(28) & 5.740(22) & 1.650(6)\\ \hline
\end{tabular}
\caption{Plaquette value, lattice spacing ($a^{-1}$) and spatial extent ($L$)
  resulting from the comparison of $w_{0}/a$ with the physical value
  quoted in the text. Errors on dimensional quantities include the
  systematic uncertainties arising from the physical value of
  $w_{0}$. }
\label{tab:lat_vol}
\end{table}

The evolution of the topological charge $Q$ (measured with the GLU
package~\cite{Hudspith_code}) is illustrated in
figure~\ref{fig:topoloical charge evolution} in appendix~\ref{sec:topology}.  These
quantities are expected to couple strongly to the slowest evolving
mode in the evolution of the algorithm~\cite{Schaefer:2010hu}.  We
obtain sets of decorrelated measurements by choosing only
configurations for further processing that are separated by $N_{\mathrm{sep}}$
intermittent update steps with $N_{\mathrm{sep}}$ larger than twice its
autocorrelation time
$\tau_{int}\left(Q^{2}\right)$~\cite{Wolff:2003sm}, as detailed in
table~\ref{tab:autocorr}.

\begin{table}
\centering{} %
\begin{tabular}{|c||cccc|}
\hline $\beta$ & $\tau_{int}\left(Q_{\textrm{top}}\right)$ &
$\tau_{int}\left(Q_{\textrm{top}}^{2}\right)$ & $N_{\textrm{sep}}$ &
$N_{{\rm cnfg}}$\\ \hline $4.41$ & $15(3)$ & $10.5(1.6)$ & $100$ &
$100$\\ $4.66$ & $160(60)$ & $74(22)$ & $200$ & $100$\\ $4.89$ &
$200(100)$ & $170\left(80\right)$ & $500$ & $111$\\ $5.20$ &
$28000\left(13000\right)$ & $12000\left(4000\right)$ & $40000$ & $36$
\\ \hline
\end{tabular}
\caption{Autocorrelation time of topological charge
  ($\tau_{int}\left(Q_{\textrm{top}}\right)$) and of charge squared
  ($\tau_{int}\left(Q_{\textrm{top}}^{2}\right)$) in units of sweep
  steps; number of sweeps separating each configuration included in
  the measured ensemble ($N_{\mathrm{sep}}$), and total number of gauge
  configurations considered.}
\label{tab:autocorr} 
\end{table}

\subsection{Observables}\label{sec:Observables}
The pseudoscalar decay constant $f_{X}$ is defined as the matrix
element of the conserved MDWF axial vector
current~\cite{Boyle:2014hxa} between a pseudoscalar meson state $X$
and the vacuum,
\[
\left\langle
0\right|\mathcal{A}_{0}\left|X\left(\mathbf{p}\right)\right\rangle
=E_X(\mathbf{ p}) f_X\,.
\]
We determine the decay constant $f_X$ and the energy $E_X(\mathbf{p})$
of the pseudoscalar state $X$ from fits to the time dependence of
Euclidean QCD two-point correlation functions projected onto momentum
$\mathbf{p}$, 
\begin{equation}
C_{MN}^{s_1,s_2}(t) \equiv \sum_{\mathbf{ x},\mathbf{
    y}}e^{i\mathbf{p}(\mathbf{x}-\mathbf{y})} \langle \,O^{s_1}_M(t,\mathbf{y})\,
\left(O^{s_1}_{N}(0,\mathbf{x})\right)^\dagger\,\rangle \stackrel{{\rm
    large}\, t}{= } \frac{Z^{s_2}_M(\mathbf{p})\left( Z^{s_1}_N(\mathbf{p}
  )\right)^\ast}{2E(\mathbf{p})} \left(e^{-E(\mathbf{p})t}\pm e^{-E(\mathbf{p}
  )(T-t)}\right)\, .
\label{eq:twopt}
\end{equation}
The operator $O^{s_i}_M$ is an interpolating operator with the quantum
numbers of the meson, i.e. $O^s_M= \bar q_2\,\omega_s\, \Gamma_M
q_1$\,, where we consider the pseudoscalar case $\Gamma_P=\gamma_5$
and the axial vector case $\Gamma_A=\gamma_0\gamma_5$,
respectively. The superscript $s$ indicates the smearing type induced
via the spacial smearing kernel $\omega$, which in the simulations
presented here is either local ($s = L$, $\omega(\mathbf{x}, \mathbf{y}) =
\delta_{\mathbf{x},\mathbf{y}}$) or Gaussian via Jacobi
iteration~\cite{Gusken:1989ad,Alexandrou:1990dq,Allton:1993wc} (see
table~\ref{tab:quark_mass_smrad} for our choice of smearing radii).
The constants $Z^{s_i}_M$ are defined by $Z^{s_i}_{M}=\langle
X(\mathbf{p})\,|\,\left( O^{s_i}_{M}\right)^\dagger\,|\,0\,\rangle$
where $X$ is the corresponding meson state.

The fits leading to the extraction of masses and decay constants are
multi-channel fits to combinations of the two-point correlation
functions $C_{AA}$, $C_{AP}$, $C_{PA}$ and $C_{PP}$.  We note the
relation between the conserved MDWF axial
current~\cite{Boyle:2014hxa,Blum:2014tka} and the renormalised local
axial current $\mathcal{A}_0=\mathcal{Z_A} A_0$, where $\mathcal{Z_A}$
is the axial vector current renormalisation constant.

A further quantity that we wish to monitor during our simulations is
the {\it residual quark mass} $am_{\rm res}$~\cite{Boyle:2014hxa},
which provides an estimate of residual chiral symmetry breaking in the
MDWF formalism. It is defined in terms of the axial Ward identity
(AWI)
\begin{equation}
\label{mobpcac}
a \Delta^-_\mu \langle (\bar \psi \gamma_5 \psi)(x)| {\cal A}_\mu(y) \rangle = \langle (\bar \psi \gamma_5 \psi)(x)|2am P(y) + 2 J_{5q}(y) \rangle \,,
\end{equation}
where $\Delta_\mu^-$ is the lattice backward derivative and $am$ is the bare quark mass in lattice units in the Lagrangian.
It motivates the definition
\[\label{eq:residual mass}
am_{\rm res}=\frac{\sum\limits_{\mathbf{ x}} \langle J_{5q}(x)
  P(0)\rangle}{\sum\limits_{\mathbf{ x}}\langle P(x) P(0)\rangle}\,.
\]
Here, $J_{5q}$ is the pseudoscalar density in the centre of the 5th
dimension.
We compute the correlation functions in eq.~\eqref{eq:twopt}
with two types of quark sources. The analysis of the decay constant
and the residual mass is based on $\mathbb{Z}_2$ noise sources and the
{\it one-end-trick}~\cite{Foster:1998vw,McNeile:2006bz,Boyle:2008rh}
(in this case we only consider $\mathbf{p}=\mathbf{0}$) while the analysis of
the dispersion relation is based on point source data.
The computation of heavy quark propagators by means of conjugate
gradient type algorithms can be affected by round-off
errors~\cite{Juttner:2005ks}.  We monitor proper convergence during
the computation of the quark propagators by checking that the desired
solver residual is fulfilled on all time slices using the \emph{time
  slice residual}~\cite{Juttner:2005ks} defined as
\begin{equation}
r_{t}={\rm
  Max}_{t}\frac{\left|D\psi-\eta\right|_{t}}{\left|\psi\right|_{t}}\,,\label{eq:timesliceresidual}
\end{equation}
where $\left|x\right|_{t}$ is the norm of the vector $x$ restricted to
time slice $t$.

We determined statistical errors using the bootstrap method with 500
samples.

\begin{table}
\centering{}%
\begin{tabular}{|cc||cc|c|lll|}
\hline $\beta$ & $L/a$ & $r_{sm}^\mathrm{P}$ & $r_{sm}^{\mathbb{Z}_2}$
& $am_{s}^{\mathrm{phys}}$ & \multicolumn{3}{c|}{$am_h$}\\ & & & &
&start&step&stop \\ \hline 4.41 & 16 & 2.8 & 4.5 & 0.03455(63)
&0.1&0.05&0.4 \\ 4.66 & 24 & 4.0 & 6.0 & 0.02416(36)
&0.066&0.033&0.396\\ 4.89 & 32 & 8.8 & 7.5 & 0.01805(33)
&0.07&0.04&0.39\\ 5.20 & 48 & 11.7 & 11.7& 0.01145(31)
&0.04&0.04&0.28\\ \hline
\end{tabular}
\caption{Simulated strange and heavy input quark masses $am_h$ and the choices of
  smearing radii for heavy quark masses. The simulated bare quark
  masses are quoted in lattice units for the MDWF action.
  The heavy quark masses starting
  from ``start'' with a step of ``step'' and ending at ``end'' are
  simulated.  $r_{sm}^\mathrm{P}$ and $r_{sm}^{\mathbb{Z}_2}$ refer to
  the choice of the smearing parameter for the Gaussian smearing of
  the source/sink of the propagators for the point and $\mathbb{Z}_2$
  noise sources, respectively. For the Gaussian smearing we employed 400
  Jacobi iterations. All measurements are carried out with MDWF with
  parameters $L_s = 12$.
\label{tab:quark_mass_smrad}}\label{tab:heavy quark masses}
\end{table}

\section{Tuning MDWF for charm\label{sec:Tuning-DWF-for}}
%
In this section we present results for the $am_h$ and $M_5$ dependence
of the heavy-heavy meson decay constant $f_{hh}$ and the residual
mass $am_{\rm res}$.

\subsection{\texorpdfstring{$M_5$}{M5} dependence}\label{sub:M5_dependence}
\begin{figure}
\centering{}
\includegraphics[width=0.45\textwidth]{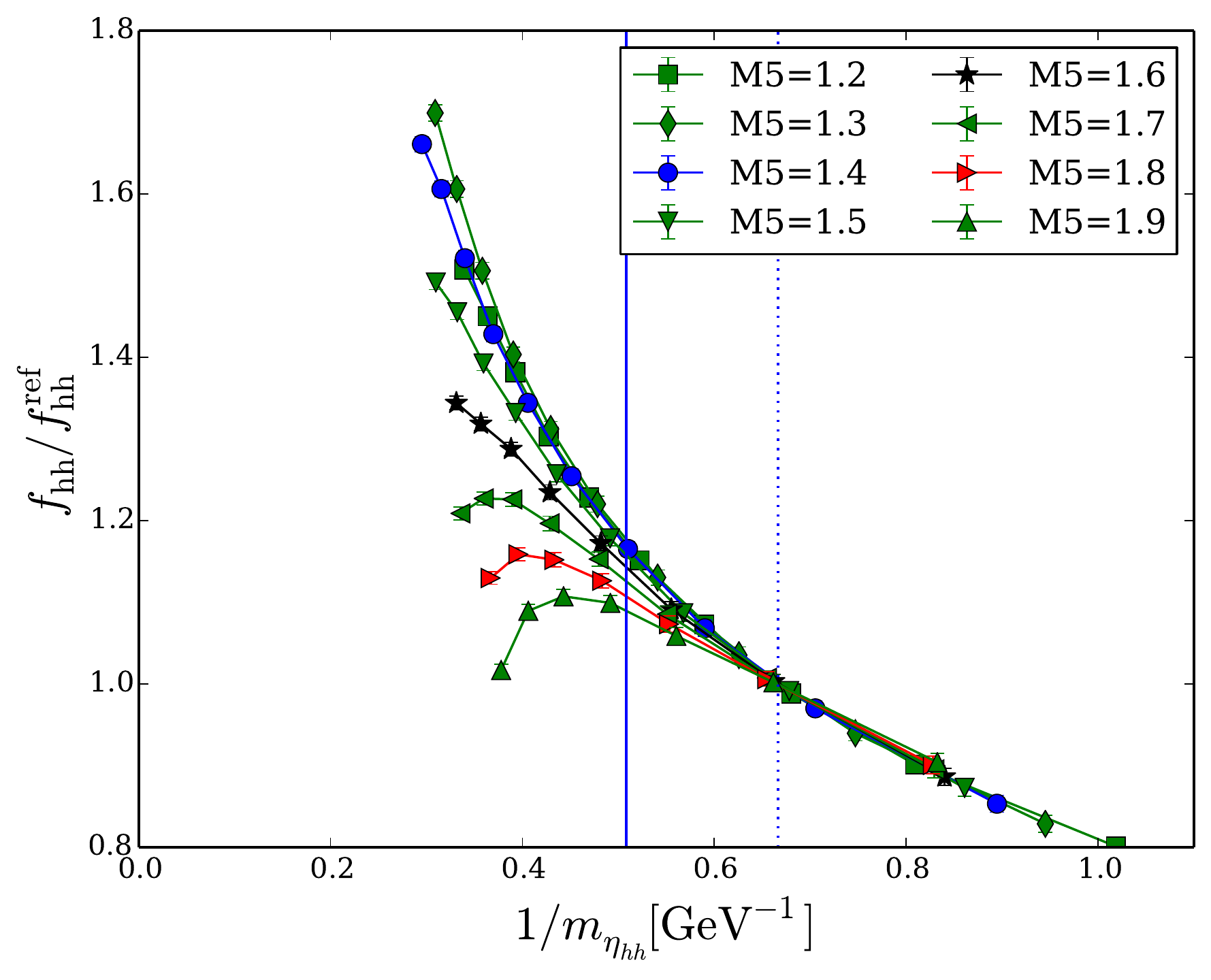}
\includegraphics[width=0.45\textwidth]{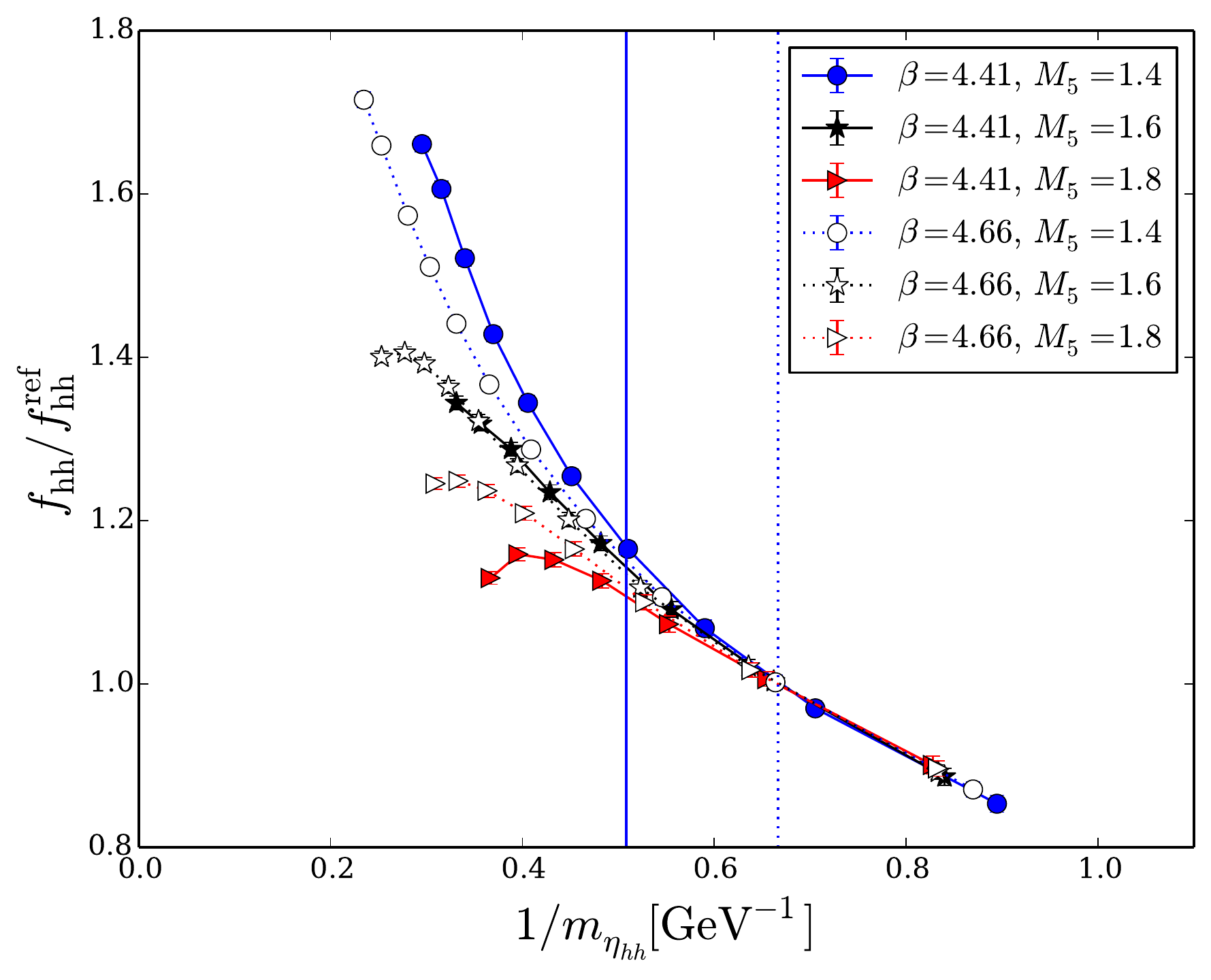}
\caption{\uline{Left}: decay constant $f_{hh}$ for heavy-heavy
  pseudoscalar mesons as a function of the inverse pseudoscalar mass
  $m_{hh}$, for different values of $M_{5}$ on the coarsest
  ensemble. The data is normalised at
  $m_{hh}^{\mathrm{norm}}=1.5\mathrm{\,GeV}$ to remove the
  multiplicative renormalisation constant. The vertical lines
  correspond to $m_{\eta_{c}}$ and to $m_{hh}^{norm}$. \uline{Right}:
  overlay of the results obtained at two different lattice spacings
  for three values of $M_{5}$.}
\label{fig:kink} 
\end{figure}

The left hand panel of figure~\ref{fig:kink} shows the dependence of
the heavy-heavy decay constant on the heavy-heavy inverse pseudoscalar
mass $m_{hh}$ observed on the coarsest ($\beta=4.41$) ensemble.  We
normalise the results for a given $M_5$ by the value of the decay
constant at $m_{hh}=1.5$\,GeV as obtained from a polynomial
interpolation.  For small values of $m_{hh}$ the decay constant shows
little dependence on the value of $M_{5}$, but as $m_{hh}$ is
increased a strong dependence is observed.

The right hand panel of figure~\ref{fig:kink} shows the same data for
$M_5=1.4,\,1.6$ and 1.8 together with the corresponding results on
the finer $\beta=4.66$ ensemble. For $M_5=1.6$ the results from the
$\beta=4.41$ and $\beta=4.66$ align almost perfectly. This provides a
first indication that for this choice of $M_5$ cutoff
effects are small. Other choices of $M_5$ would offer viable alternatives
but with more pronounced cutoff effects.

\subsection{Residual mass}
\begin{figure}
\centering{}%
\begin{tabular}{cc}
\includegraphics[width=0.49\textwidth]{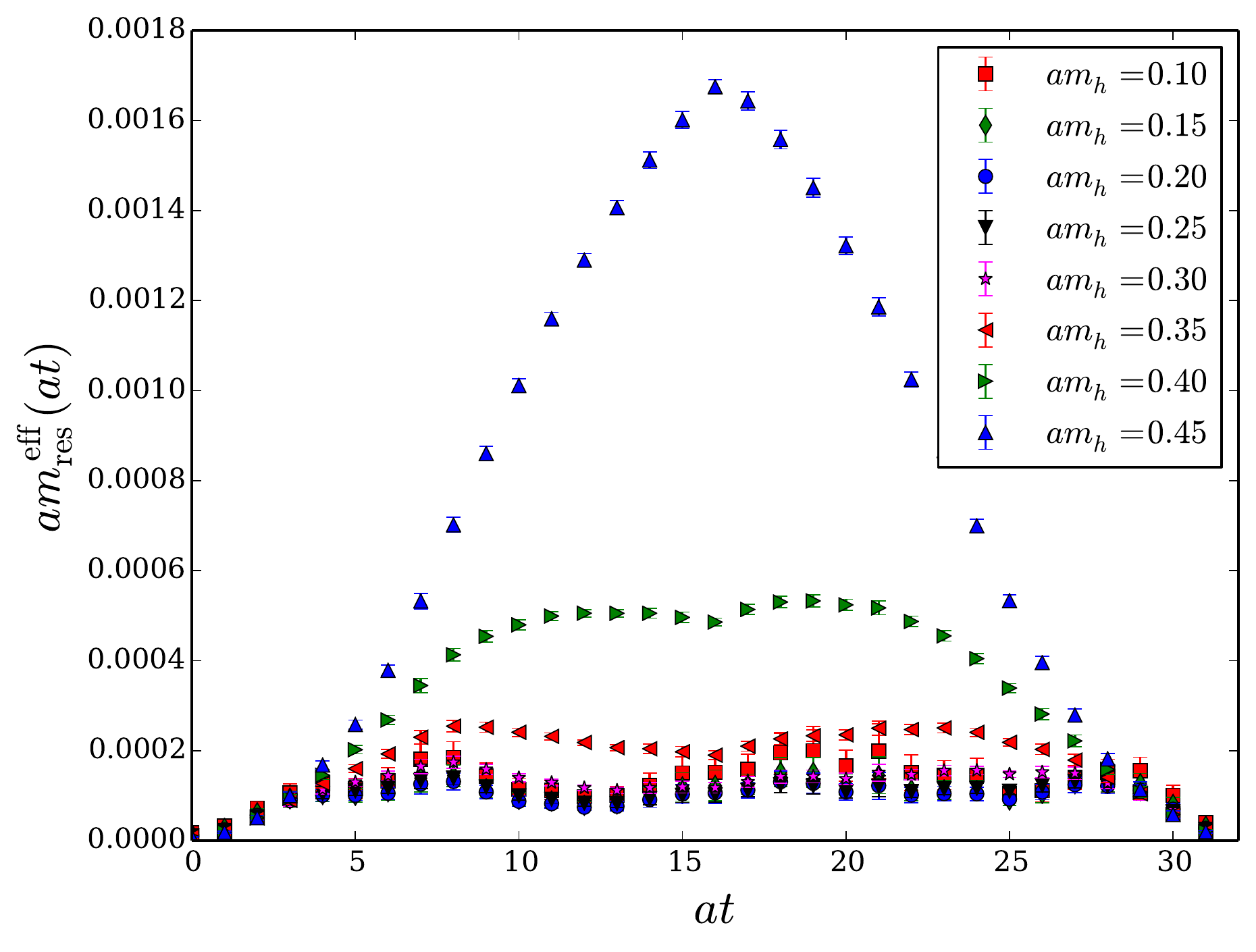} &
\includegraphics[width=0.49\textwidth]{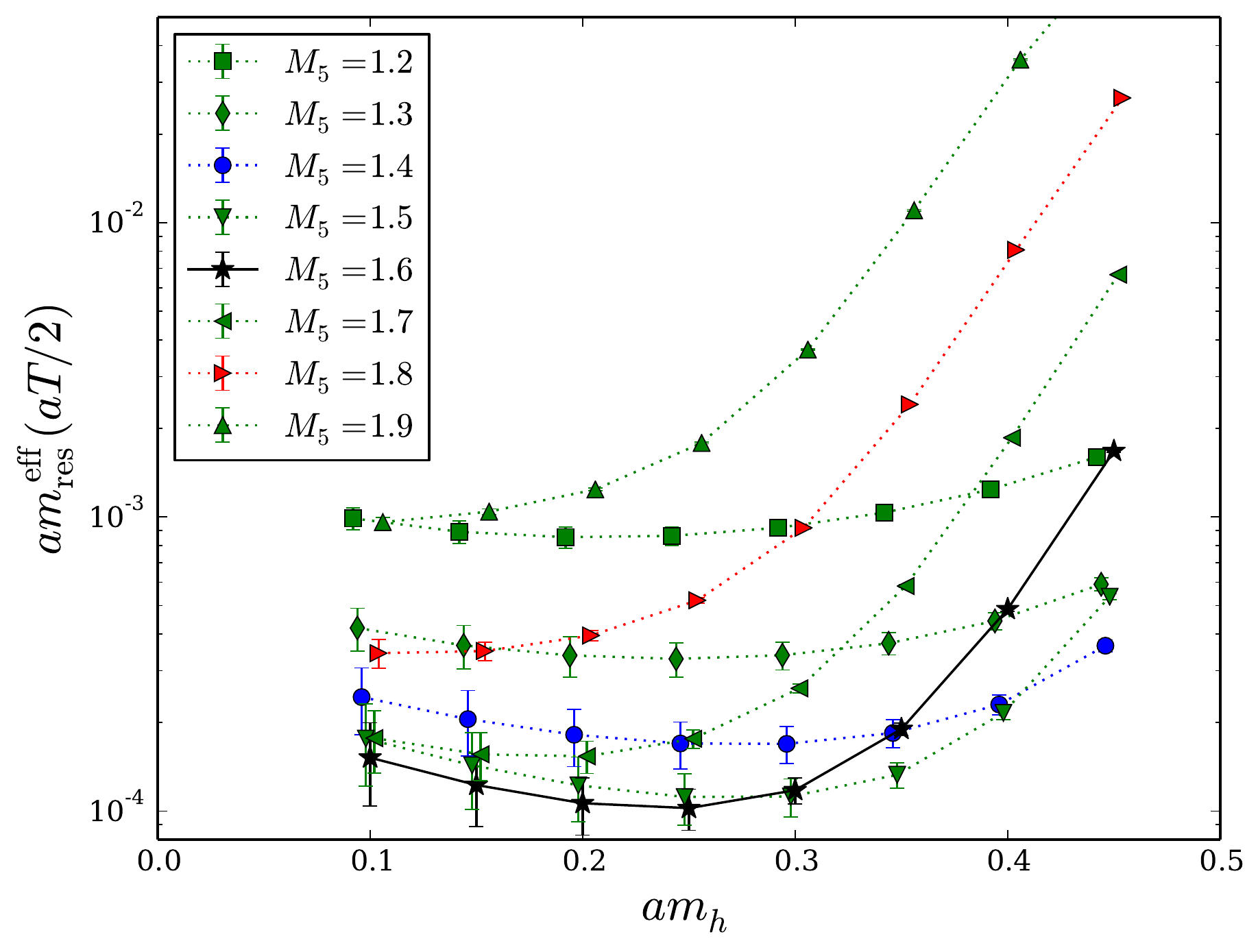}\\
\end{tabular}
\caption{\uline{Left}: behaviour of the \emph{effective} residual mass
  $am_{\mathrm{res}}^{\mathrm{eff}}$ as a function of time on our
  coarsest ensemble, for $M_{5}=1.6$. \uline{Right}: residual mass
  determined at $t=T/2$ as a function of the bare quark mass for
  several values of $M_{5}$.}
\label{fig:mres}
\end{figure}

Next we quantify how the residual chiral symmetry breaking is affected
by $M_5$ by observing the response of the size of the residual mass to
variations in $am_h$ and $M_5$.
In the left panel of figure~\ref{fig:mres} we show the ratio of
correlation functions eq.~\eqref{eq:residual mass} from which we
determine $am_{\rm res}$ as a function of time for several values of
the quark mass at $M_{5}=1.6$. Note that for large $t$ the time
dependence in ratio eq.~\eqref{eq:residual mass} is expected to cancel
between the numerator and denominator. While the expected (constant)
behaviour in time is observed for small quark masses, this is
strikingly not the case for values of $am_{h}^{bare}\gtrsim 0.4$.

In these cases it is difficult to interpret the operator $J_{5q}$’s
matrix element as a constant, residual additive mass correction in the
chiral Ward identity.  The effect is of course rather small compared
to the explicit chiral symmetry breaking, but there is a risk that the
physical modes no longer remain bound to the walls of the fifth
dimension in this large mass limit. To be more quantitative, we define
$am_{\mathrm{res}}(t=T/2)$ as the value of this correlator ratio in
the (temporal) middle of the lattice. Note, however, that above 
$a m_h\approx 0.4$ the meaning of $a m_{\rm res}$ as a unique measure of
residual chiral symmetry breaking is no longer clear, only
indicative. The right hand panel in figure \ref{fig:mres} shows
$am_{\rm res}(t=T/2)$ as a function of the quark mass. We observe the
same qualitative behaviour for all values of $M_5$: as the input quark
mass is increased beyond $a m_h \approx 0.4$ the residual mass $am_{\rm
  res} (t=T/2)$ starts to increase drastically. Although this quantity
is $L_s$ dependent, it is likely unsafe to use domain wall fermions at
masses where the physical modes become unbound from the walls and the
matrix elements of $J_{5q}$ have such non-trivial behaviour.  The
impact on 4$d$ observables will be studied later in this paper.
%

\subsection{Locality of the effective 4$d$ Dirac operator}
Given the above observation indicating the reduced binding of surface states of MDWF above
$am_h\approx 0.4$, a further concern one might have is that we should check the locality
property of the corresponding effective 4$d$ MDWF Dirac operator. 
The connection of the 5$d$ MDWF operator $D^5_{MDWF}$ defined in eq.~(\ref{eq:MDWFop})
to a four dimensional effective theory is well established in the literature, 
\cite{Borici:1999zw,Borici:1999da,Brower:2004xi,Brower:2005qw,Brower:2012vk,Kennedy:2006ax,Blum:2014tka}.
We identify $D_{ov}$ as an approximation to the overlap operator with 
approximate sign function
\begin{equation}
\epsilon(H_M) = \frac{ (1+ H_M)^{L_s}-(1 - H_M)^{L_s}}
                     { (1+ H_M)^{L_s}+(1 - H_M)^{L_s}}\,,
\label{eq:sign_approx}
\end{equation}
where the M\"obius kernel is
\begin{equation}
H_M = \gamma_5\frac{(b+c) D_W}{2+(b-c)D_W}\,.
\end{equation}
The transfer matrix in the fifth dimension can be identified as
\begin{equation}
\begin{array}{ccc}
T^{-1} = -[ H_M - 1]^{-1} [ H_M + 1].
\end{array}
\end{equation}
The effective overlap operator may be simply found  as
\begin{eqnarray}
D_{ov} &=& \left[{\cal P}^{-1} D^5_{MDWF}(am=1)^{-1}  D^5_{MDWF}(am) {\cal P}\right]_{11}\,\\
       &=&\left[ \frac{1+am}{2} + \frac{1-am}{2}\gamma_5 \frac{T^{-L_s}-1}{ T^{-L_s}+1} \right],\label{eq:Dov}
\end{eqnarray}
where this is known to reduce to the standard overlap formalism in the limit $L_s\to\infty$ and when $b=c$,
and the projection matrix ${\cal P}$ is
\begin{eqnarray}
{\cal P} &=& \left( 
\begin{array}{ccccc}
P_- & P_+     & 0      & \ldots & 0 \\
0   & \ddots & \ddots & 0 & \vdots \\
\vdots& 0 & \ddots & \ddots & 0 \\
0    & \ldots & 0 & \ddots & P_+ \\
P_+  & 0 & \ldots & 0 & P_-
\end{array}
\right).
\end{eqnarray}
Following eq.~\eqref{eq:Dov}, we may place the mass dependence of $D_{ov}(am)$ at non zero mass in the following form:
\begin{eqnarray}
D_{ov}(am)   &=&\left[ \frac{1+am}{2} + \frac{1-am}{2}\gamma_5 \frac{T^{-L_s}-1}{ T^{-L_s}+1} \right]\\
            &=& am + (1-am) D_{ov}(0)\label{eq:Dovm0} \\
            &=& (1-am) \left[ \frac{am}{1-am} + D_{ov}(0) \right]\label{eq:Dovm}.
\end{eqnarray}
We see that the kinetic term in the four dimensional effective action
should remain unaltered as the mass is changed up to a trivial
rescaling factor $(1-am)$ affecting the surface field renormalisation.
The induced overlap bare mass is therefore better interpreted as the
combination $\frac{am}{1-am}$, which of course varies non-linearly and
diverges as we take the domain wall mass towards the Pauli-Villars
mass of unity.  The exponential locality~\cite{Hernandez:1998et} is
fully encoded in the massless operator, and is independent of the
quark mass. So, from this perspective there should be no locality
issues as we take the mass large, since the kinetic term is trivially
rescaled compared to the light mass case.

We demonstrate this with a second use of eq.~\eqref{eq:Dov}.
The effective operator may be constructed by the simple
application of the inverse of the Pauli Villars operator.
Following the methodology of
ref.~\cite{Hernandez:1998et} we now study the locality properties of this operator.

We start by defining a point source $\xi$,\texttt{ }
\[
\xi_{\alpha,a}\left(x\right)=\begin{cases} 1 & x=y,\ \alpha=a=0\,({\rm spin, colour})\\ 0 & {\rm otherwise},
\end{cases}
\]
where $y$ is the source location,
and $\psi$ is the result of the multiplication of the effective 4$d$ Dirac operator
with $\xi$,
\[
\psi=D_{ov}\xi\,.
\]
We say $D_{ov}$ is strictly local (or ``ultralocal'') if the only non-zero
contributions to $\left(D_{ov}\xi\right)\left(x\right)$ come from a finite
set of terms $D_{ov}\left(x,y\right)\xi\left(y\right)$ with $y$ in the
vicinity of $x$\textbf{~}\cite{Hernandez:1998et}.

We collect all lattice points $\left\{ x\right\} _{r}$ separated by
$r$ hoppings from the origin, such that $x\in\left\{ x\right\} _{r}$
if $\left|x\right|_{1}=r$. Here $\left|x\right|_{1}$ is the ``taxi
driver'' (or ``Manhattan'') norm of $x$, defined by
\begin{equation}
  \left|x\right|_{1}=\sum_{\mu}\min\left\{
    \left|x_{\mu}\right|,\,\left|N_{\mu}-x_{\mu}\right|\right\} \,,
\end{equation}
where
$N_{\mu}$ is the number of lattice sites along the $\mu$ axis.
This definition accounts for the periodicity of the lattice.
Finally, for each value of $r$ we define the maximum of the norm of
$\psi$ at the set of points $\left\{ x\right\} _{r}$:
\begin{equation}\label{eq:fdef}
f\left(r\right)=\max\left\{ \left|\psi\left(x\right)\right|\forall\,
x\in\left\{ x\right\} _{r}\right\} \,.
\end{equation}
In the following we will study $f\left(r\right)$ for values of the
bare heavy quark mass {in lattice units} of $am_h=0.1$ and $am_h=0.5$
with $M_{5}=1.6$ on our $\beta=4.41$, $\beta=4.66$ and $\beta=4.89$
ensembles.

\begin{figure}
\noindent \begin{centering}
  \includegraphics[scale=0.45]{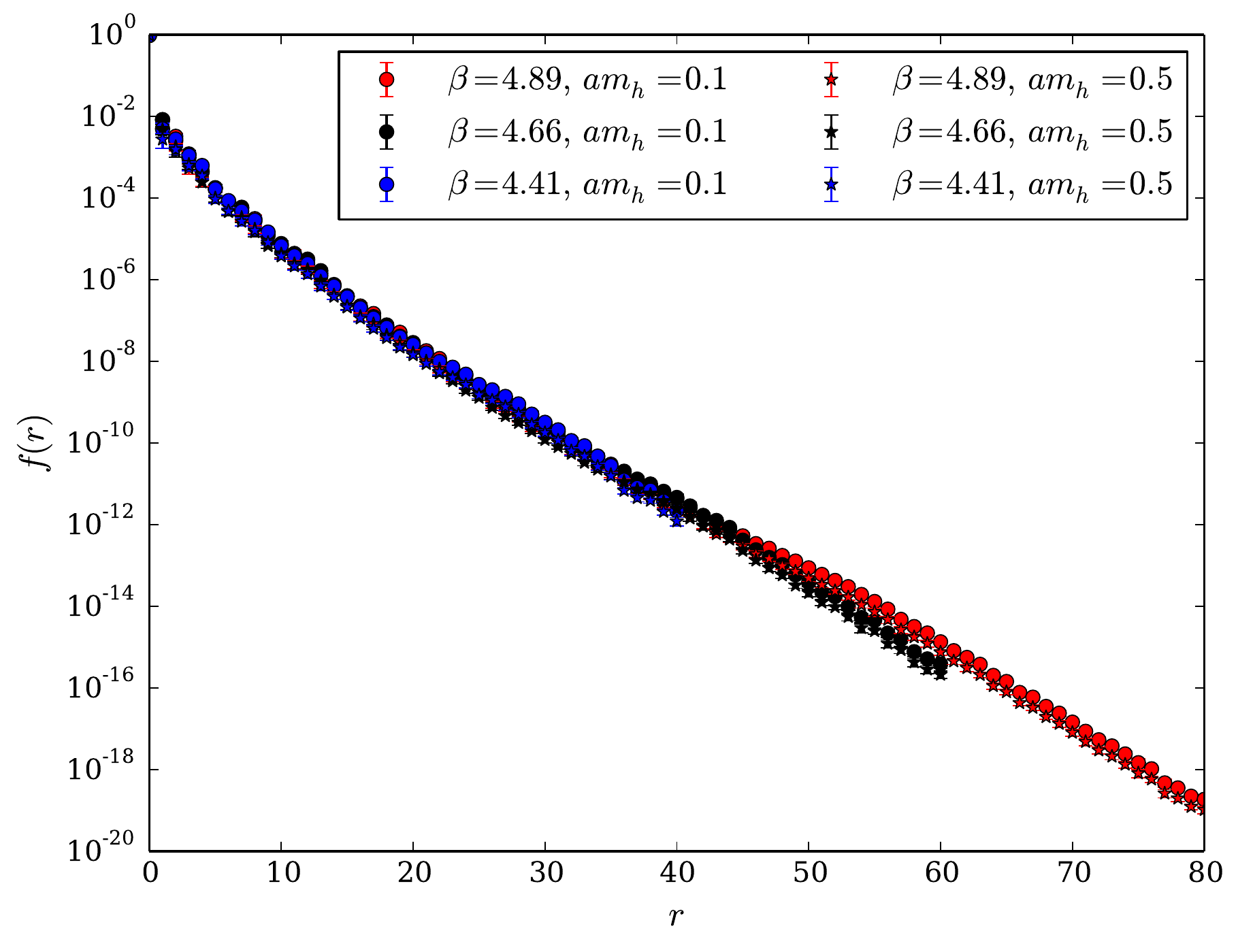}
  \par\end{centering}

\caption{\label{fig:Localisation-curves-free}Localisation function
  (with logarithmic y-scale) for the effective MDWF operator at two
  bare quark masses on the three coarser quenched ensembles.}
\end{figure}

In figure~\ref{fig:Localisation-curves-free} we show
the function $f\left(r\right)$ for two bare quark masses on all three
ensembles.  As expected, we observe that the slope of $f\left(r\right)$ is
independent of the bare quark mass as well as of the lattice spacing,
indicating that locality is recovered in the continuum limit.

We can make a more quantitative statement for the mass independence of
the locality of $D_{ov}(am)$: motivated by eq.~(\ref{eq:Dovm0}) we
define the function $\tilde{f}$:
\begin{equation}\label{eq:fdef}
\tilde{f}_{m}\left(r\right)=\max\left\{ \left|\psi\left(x\right) - am\,\xi(x) \right|\forall\,
x\in\left\{ x\right\} _{r}\right\} \,,
\end{equation}
where we have introduced a term to subtract the additive mass term in
eq.~(\ref{eq:Dovm0}). We can then define the ratio
\begin{equation}
R(r) = \frac{\tilde{f}_{m_1}(r)(1-am_2)}{\tilde{f}_{m_2}(r)(1-am_1)}\,,
\end{equation}
where the subscripts indicate the  bare quark masses at which the function $f$
was evaluated ($am_1=0.1$ and $am_2=0.5$). According to eq.~(\ref{eq:Dovm0}),
we expect $R(r)=1$, which is confirmed by our data to the level of 
arithmetic precision used in the computation.
This provides a strong consistency check of our setup and our understanding 
of the locality of the MDWF operator.

\section{Continuum limit of the decay constant and the dispersion relation\label{sec:Continuum-limit}}
The results in the previous section provide the first evidence for a
region in parameter space where MDWF can be used as a suitable
discretisation for heavy quarks.
To further substantiate this picture we now fix $M_5=1.6$ and study
the continuum scaling of a basic heavy-strange pseudoscalar meson
matrix element, the decay constant, and the corresponding dispersion
relation, as a function of the mass of the heavy quark.

\subsection{Choice of strange and heavy quark masses}
We study the continuum limit along lines of constant strange and heavy
quark mass.
We fix the $s$-quark by considering a fictitious meson $\eta_{s}$
composed of two different quarks, $s$ and $s^{\prime}$, of degenerate
mass $m_{s}$. This meson differs from the physical
$\eta-\eta^{\prime}$ mesons by quark-disconnected Wick contractions.
We tuned the strange quark mass to its ``physical value'', by imposing
at each lattice spacing the mass of the simulated $\eta_{s}$ meson to
reproduce
$m_{\eta_{s}}=0.6858(40)\,\mathrm{GeV}$~\cite{Davies:2010ip}. This sets
a common renormalised strange quark mass on all the ensembles.  In
table~\ref{tab:quark_mass_smrad} we report on the values of the
corresponding bare strange quark mass
and on our choices for the simulated heavy quark masses. 

\subsection{Decay constants for heavy-strange mesons}
We consider the renormalised ratio
\[
R_{sh}=\frac{f_{sh}\sqrt{m_{sh}}}{f_{sh}^{norm}\sqrt{m_{sh}^{norm}}}\,,
\]
where we introduce $f_{sh}^{norm}\sqrt{m_{sh}^{norm}}$, interpolated
to $m_{sh}=1$\,GeV, to cancel the axial current renormalisation constant. We also
include in both the numerator and denominator a factor of $\sqrt{m_{sh}}$ to make
both of these quantities individually finite in the limit $am_h\to\infty$.

\begin{figure}
\begin{centering}
\includegraphics[width=0.45\textwidth]{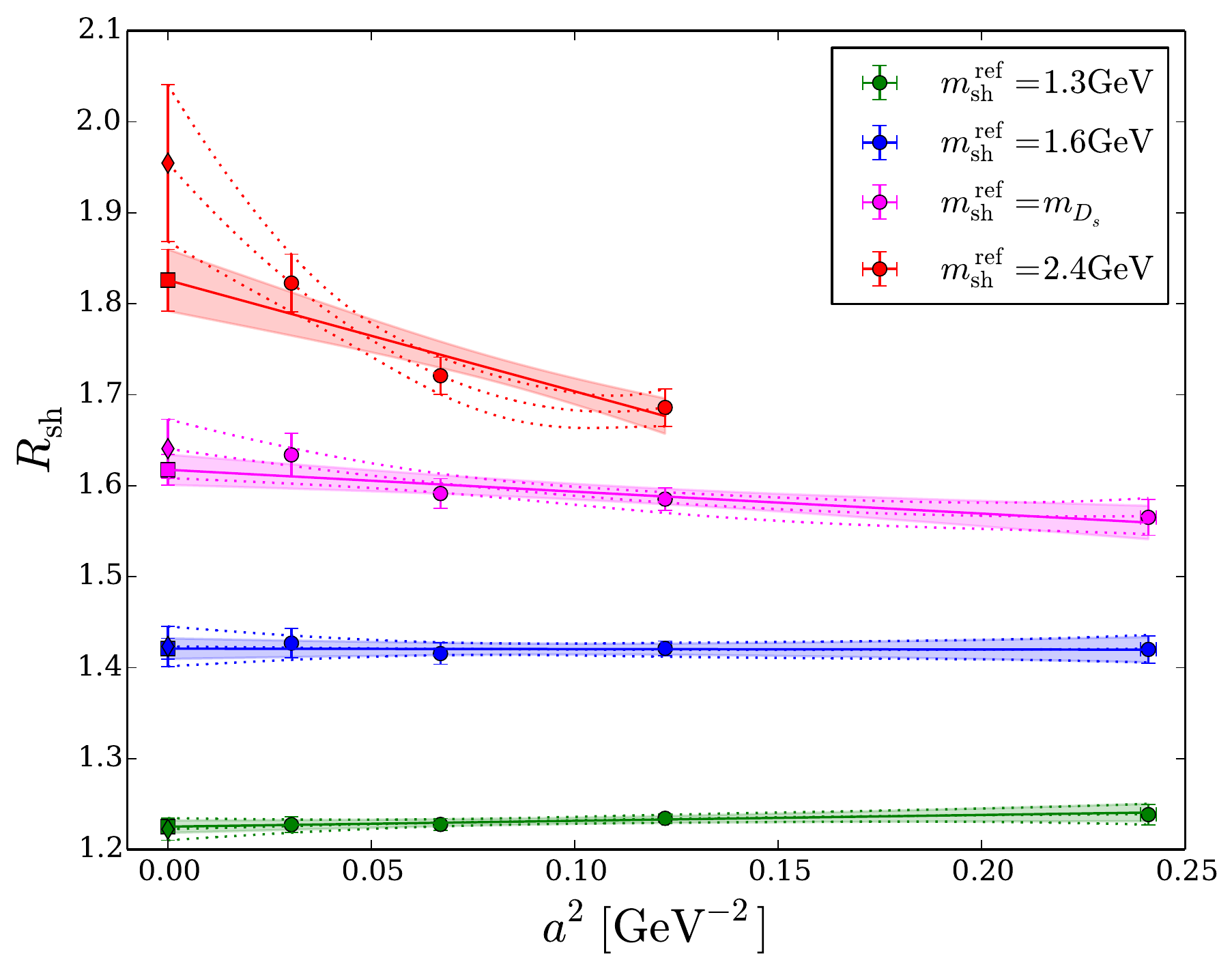}
\par\end{centering}
\caption{Continuum limit of the ratio of heavy-strange decay constants
  at different reference pseudoscalar masses with linear (dashed
  shaded error band, square-symbols) and quadratic (dotted lines,
  diamond symbols) polynomials in $a^{2}$.\label{fig:cont}}
\end{figure}

We interpolate $R_{sh}$ to the reference pseudoscalar masses 1.3, 1.6,
$m_{D_s}=1.9685$~\cite{Agashe:2014kda} and 2.4\,GeV on all ensembles. To
fulfil the constraint $am_h\leq 0.4$ we are forced to drop the
coarsest lattice spacing for the heaviest mass considered. A first
visual inspection (see figure~\ref{fig:cont}) suggests the absence of
cutoff effects beyond $O(a^2)$. Moreover, cutoff effects are observed
to be very mild for the choice $M_5=1.6$, in agreement with the
observation made in section~\ref{sec:Tuning-DWF-for}.

To obtain a more quantitative understanding we perform
continuum limit extrapolations by considering two different fit
ans\"atze, namely
\begin{eqnarray}
R_{1}\left(a\right) & \equiv & R_{a=0}+D_{2}a^{2}\,,\nonumber
\\ R_{2}\left(a\right) & \equiv &
R_{a=0}+E_{2}a^{2}+E_{4}a^{4}\,.\label{rats}
\end{eqnarray}
The results are illustrated in figure~\ref{fig:cont} as solid and dashed
lines with error bands, respectively, and the resulting fit
coefficients are listed in
table~\ref{tab:CLdecay}.

For the two lightest reference masses, 1.3 and 1.6\,GeV, the slope of
the continuum limit is compatible with zero. For higher masses the
continuum limit starts exhibiting a significant slope.  In fact, the
dimensionless term $D_{2}a^{2}/R\left(a=0\right)$, which indicates the
fractional amount of discretisation errors, is around 2\%
for the physical $D_{s}$ meson on the coarsest ensemble
($a^{-1}\approx2$\,GeV), and of $\mathcal{O}\left(1\%\right)$ on
the next finest one ($a^{-1}\approx2.9$\,GeV).  At the level of statistical
precision achieved here the fits reveal only a very mild sensitivity to
higher order ($O(a^4)$) coefficients: $E_4$ is compatible with
zero within one standard deviation.
\begin{table}
{\small
\centering{}%
\begin{tabular}{|c||cccc|ccccc|}
\hline $m_{sh}^{\mathrm{ref}}[\mathrm{GeV}]$ & $R_{a=0}$ & $D_{2}[\mathrm{GeV}^{2}]$ & $\chi^{2}/\mathrm{dof}$ & $p$ & $R_{a=0}$ & $E_{2}[\mathrm{GeV}^{2}]$ & $E_{4}[\mathrm{GeV}^{4}]$ & $\chi^{2}/\mathrm{dof}$ & $p$\\ \hline
1.3       & 1.225(06) & 0.06(06) & 0.10  & 0.95 & 1.222(12) & 0.12(19) & -0.21(71) & 0.12 & 0.73\\
1.6       & 1.421(11) & 0.00(09) & 0.18  & 0.91 & 1.423(21) & -0.05(34) & 0.2(1.2) & 0.34 & 0.56\\
$m_{D_{s}}$ & 1.618(16) & -0.24(12) & 0.75 & 0.69 & 1.641(32) & -0.67(51) & 1.5(1.7) &0.84 & 0.36\\
2.4       & 1.826(33) & -1.22(37) & 2.64 & 0.10 & 1.955(86) & -5.0(2.4) & 23(15) & - & -\\ \hline
\end{tabular}}
\caption{Results of the continuum limit extrapolation for the
  heavy-strange decay constants. The first block summarises the
  results for the linear extrapolation in $a^2$, the second block the
  quadratic extrapolation in $a^2$. We also show corresponding
  results for the $\chi^2/{\rm dof}$ and $p$-values. \label{tab:CLdecay}}
\end{table}

\subsection{Dispersion relation}
On the lattice, the continuum dispersion relation for pseudoscalar
mesons
\begin{equation}
E\left(m,\mathbf{p}\right)=\sqrt{m^{2}+\mathbf{p}^{2}}\,,\label{eq:contdisprel}
\end{equation}
is modified: all even powers of the lattice spacing with $p$-dependent
coefficients, invariant under hypercubic group transformations
(e.g. $p^{2}$, $\sum_{\mu}p_{\mu}^{2n}$...), are allowed.  Here we
investigate whether the continuum expression is correctly reproduced
after taking the continuum limit of the lattice data for the
heavy-strange meson energy at various momenta $\mathbf{p}=\frac{2\pi}{L}\mathbf{n}$. In particular, we consider the cases
$\mathbf{n}\in\left\{ \left(0,0,0\right),\,\left(1,0,0\right),\,
\left(1,1,0\right),\,\left(1,1,1\right)\right\}$.

The measured meson energies are sufficiently precise to be sensitive
to the slight mistunings in the physical volume of our ensembles
(cf. table~\ref{tab:lat_vol}). In particular, for any given
$\mathbf{n}$ the simulated lattice momenta $\mathbf{p}^\mathrm{sim}$
in physical units only agree approximately amongst the different
ensembles.

We correct for this by defining a reference volume with spatial extent
$L^\mathrm{ref}=1.648$\,fm and therefore reference momenta
$\mathbf{p}^\mathrm{ref}=\frac{2 \pi}{L^\mathrm{ref}} \mathbf{n}$.
The meson energies $E^\mathrm{sim}$ are interpolated to this by taking
advantage of the lattice dispersion relation
\begin{equation}
\sinh^{2}\left(\frac{aE}{2}\right)=\sinh^{2}\left(\frac{am}{2}\right)+\sum_{i=1}^{3}\sin^{2}\left(\frac{ap_{i}}{2}\right)\,.\label{eq:latt-disp-rel}
\end{equation}
Considering eq.~\eqref{eq:latt-disp-rel} for a meson of momentum
$\mathbf{p}^\mathrm{ref}$ on two different volumes we obtain the
interpolated energy:
\begin{equation}
E^{\mathrm{ref}}=2a^{-1}\sinh^{-1}\sqrt{\sinh^{2}\left(\frac{aE^{\mathrm{sim}}}{2}\right)-
  \sum_{i=1}^{3}\sin^{2}\left(\frac{ap_{i}^{\mathrm{sim}}}{2}\right)+\sum_{i=1}^{3}\sin^{2}\left(\frac{ap_{i}^{\mathrm{ref}}}{2}\right)}\,.\label{eq:momentum-correction}
\end{equation}

\begin{figure}
\begin{centering}
\includegraphics[width=0.6\textwidth]{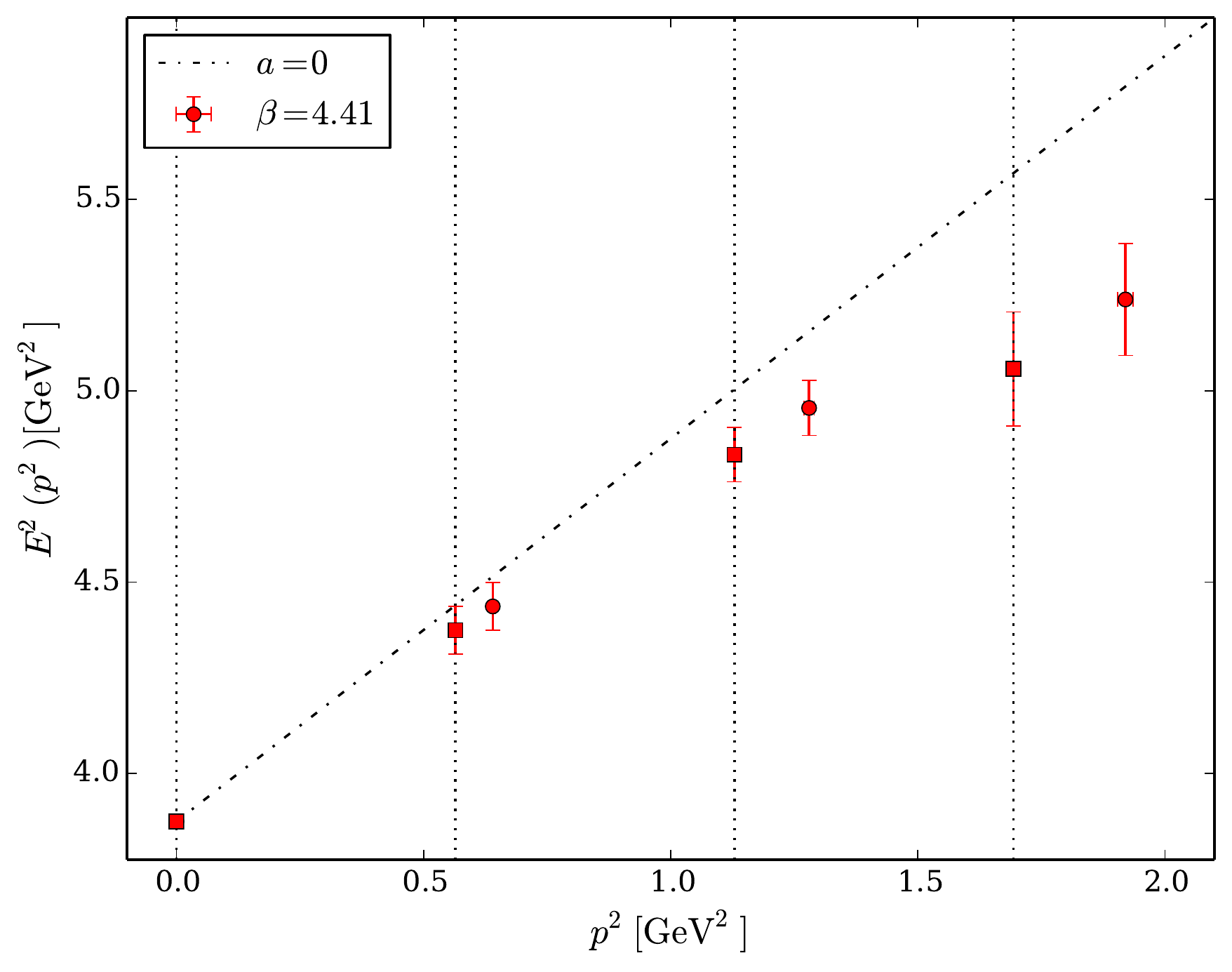}
\par\end{centering}
\caption{Interpolation to reference momenta (marked by vertical dotted
  lines) for the coarsest ensemble at the physical $D_s$ mass. The
  black dash-dotted line depicts the continuum dispersion relation,
  displayed only for reference. The closed circles show the simulated
  data on the coarsest ensemble ($\beta=4.41$) whilst the closed
  squares which lie on the vertical lines correspond to the values
  after the correction of eq.~\eqref{eq:momentum-correction} was
  applied. \label{fig:Interpolation-to-reference-momentum}}
\end{figure}

In figure~\ref{fig:Interpolation-to-reference-momentum} we show an
example of the interpolation to the chosen reference momenta for the
ensemble requiring the largest corrections
(cf. table~\ref{tab:lat_vol}) for the case of the physical-mass $D_{s}$
meson.

\begin{figure}
\begin{centering}
\begin{tabular}{cc}
\includegraphics[width=0.49\textwidth]{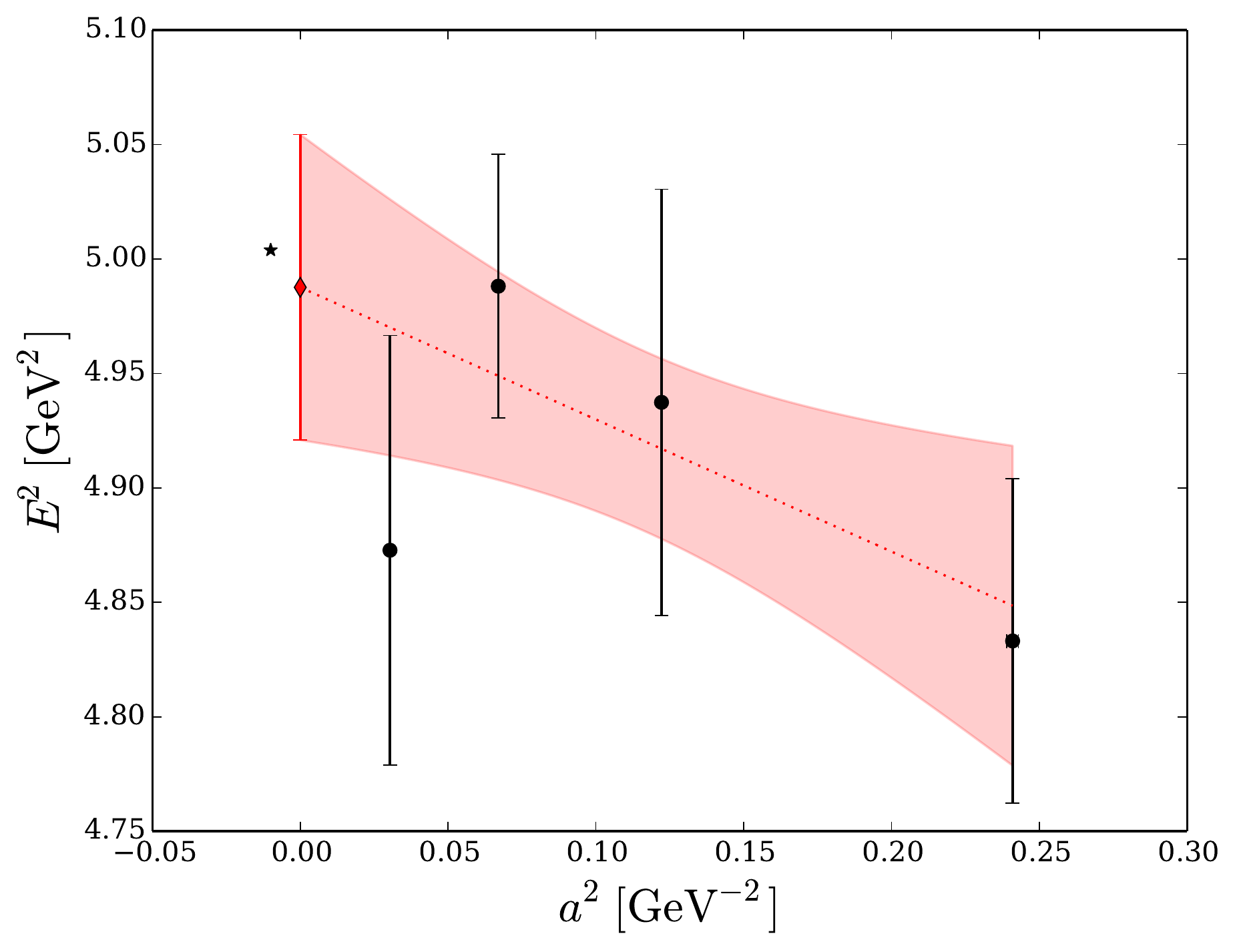}
&
\includegraphics[width=0.49\textwidth]{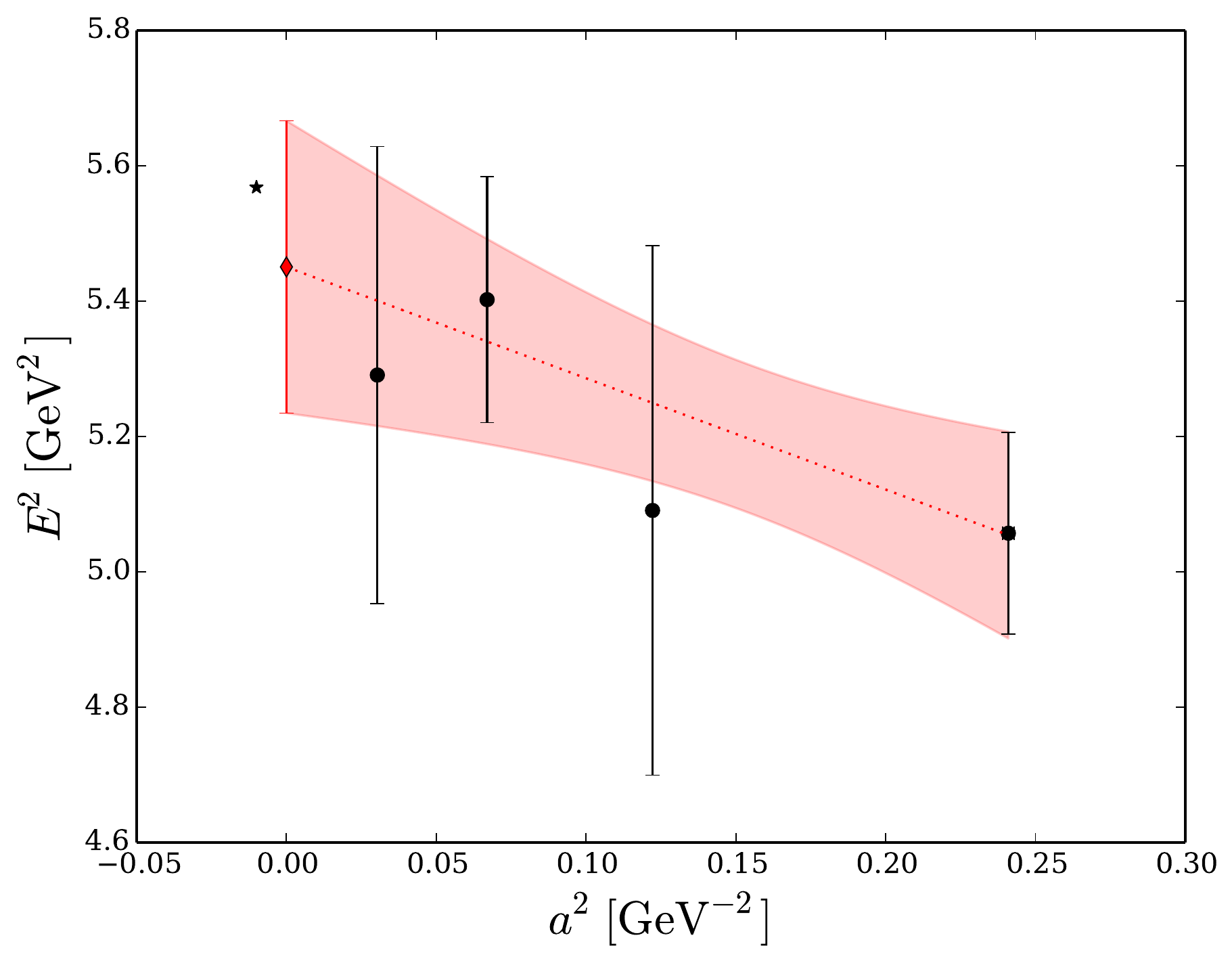}\\
\end{tabular}
\par\end{centering}

\caption{Continuum limit for momenta $(1,1,0)$ (left panel) and
  $\left(1,1,1\right)$ (right panel) for the physical-mass $D_{s}$
  meson. Black circles correspond to finite lattice spacings, red
  diamonds to the continuum limit extrapolation; the band shows the fit
  ansatz, whereas the star is the energy of the meson computed using
  eq.~\eqref{eq:contdisprel} and the meson rest
  mass. \label{fig:Continuum-Limit-momentum}}
\end{figure}

We now proceed to perform the continuum limit extrapolation of the
meson energy. In figure~\ref{fig:Continuum-Limit-momentum} we
illustrate the extrapolation of the physical $D_{s}$ meson energies
for two different momenta. In both cases the extrapolated result is
compatible with the energy predicted by the continuum dispersion
relation~\eqref{eq:contdisprel}.

\begin{figure}
\begin{centering}
\includegraphics[width=0.6\textwidth]{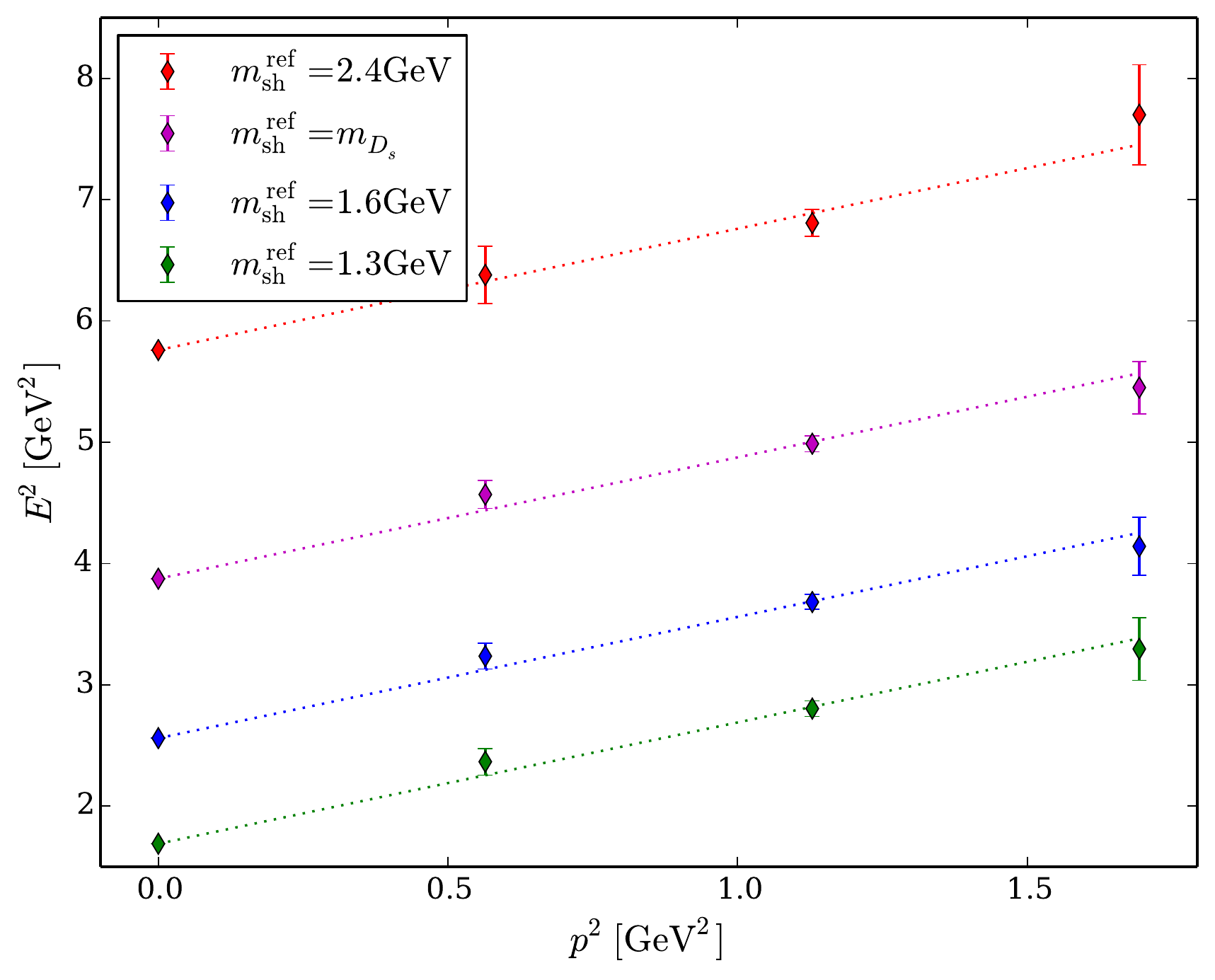}
\par\end{centering}
\caption{Continuum extrapolated results for the meson energy as a
  function of the momentum for the different reference masses. Each
  dotted line depicts the continuum dispersion relation
  eq.~\eqref{eq:contdisprel} for the associated reference mass.
  For the heaviest reference mass
  ($m_{sh}=2.4\mathrm{\,GeV}$), only the three finer ensembles enter the
  continuum limit since the heavy mass reach of the coarsest ensemble
  is not sufficient. This results in the larger
  errorbars. \label{fig:CLdisprel all}}
\end{figure}

This procedure was repeated for all momenta and reference masses. In
figure~\ref{fig:CLdisprel all} we show the results for the energies
after the continuum limit extrapolation for the different momenta and
choices of reference rest masses. The expected continuum
dispersion relation eq.~\eqref{eq:contdisprel} is recovered,
indicating a good control over the continuum limit.

\section{Conclusion\label{sec:Conclusion}}

This study is motivated by the need to explore new and alternative
ways for discretising heavy flavours in simulations of lattice QCD:
more independent predictions for heavy flavour hadronic quantities
with a solid control of systematic uncertainties are urgently
needed~\cite{Aoki:2013ldr} to make reliable predictions for SM
phenomenology.

To this end we explored the feasibility of M\"obius domain wall
fermions (MDWF) as a lattice regularisation for heavy quarks. DWF have
so far been widely used as a discretisation for the light $u$, $d$ and
$s$ quarks. Its desirable features are chiral symmetry to a
good approximation and automatic ${\cal O}(a)$-improvement.

From our simulations within quenched QCD with the tree-level improved
Symanzik gauge action we have identified a point in MDWF parameter space, 
the domain wall height $M_5=1.6$, for which
discretisation effects turn out to be particularly small.
We demonstrated that the salient features of MDWF persist for heavy
quarks as long as the bare input quark mass obeys the bound
$am_h\lesssim 0.4$.
Based on these findings we carried out a detailed scaling study of the
heavy-strange dispersion relation and decay constant. Over the
range of lattice cutoffs 2.0--5.7~GeV the observables were compatible with
a linear scaling in $a^2$. At the level of precision achieved in this
work, coefficients of $a^4$ terms were found to be almost always
compatible with zero, remaining remarkably  small even for the
heaviest quark mass (heavier than charm) simulated.

The results accumulated in this paper constitute a proof of concept
for MDWF as a powerful discretisation to study charm and heavier
quarks on current dynamical gauge field ensembles. This work
constitutes a solid basis for RBC/UKQCD's heavy MDWF phenomenology
program~\cite{Boyle:2015kyy}. Nevertheless, we are
considering ideas for how to improve the current setup, for example
by link-smearing the MDWF kernel \cite{Cho:2015ffa,Boyle:2015kyy}.

\subsection*{Acknowledgements}
The authors
are thankful for many fruitful discussions with Y.Cho,
S.Hashimoto, T. Kaneko and J. Noaki from  K.E.K. and with our colleagues
in the RBC and UKQCD collaborations.
The research leading to these results has received funding from the
European Research Council under the European Union's Seventh Framework
Programme (FP7/2007-2013) / ERC Grant agreement 279757. P.A.B. is supported in part by 
UK STFC Grants No. ST/M006530/1, ST/L000458/1, ST/K005790/1, and ST/K005804/1.
M.S. is supported by UK EPSRC Doctoral Training Centre
Grant EP/G03690X/1. 
The authors
gratefully acknowledge computing time granted through the STFC funded
DiRAC facility (grants ST/K005790/1, ST/K005804/1, ST/K000411/1,
ST/H008845/1). The authors acknowledge the use of the IRIDIS High
Performance Computing Facility, and associated support services at the
University of Southampton, in the completion of this work. 
\clearpage

\appendix

\section{Topological charge evolution}\label{sec:topology}
In figure~\ref{fig:topoloical charge evolution} we show the Monte Carlo
histories and histograms of the topological charge restricted to
the configurations on which we also determined the decay constant and
the meson energies.
  \begin{figure}
    \centering
    \includegraphics[width=.32\textwidth]{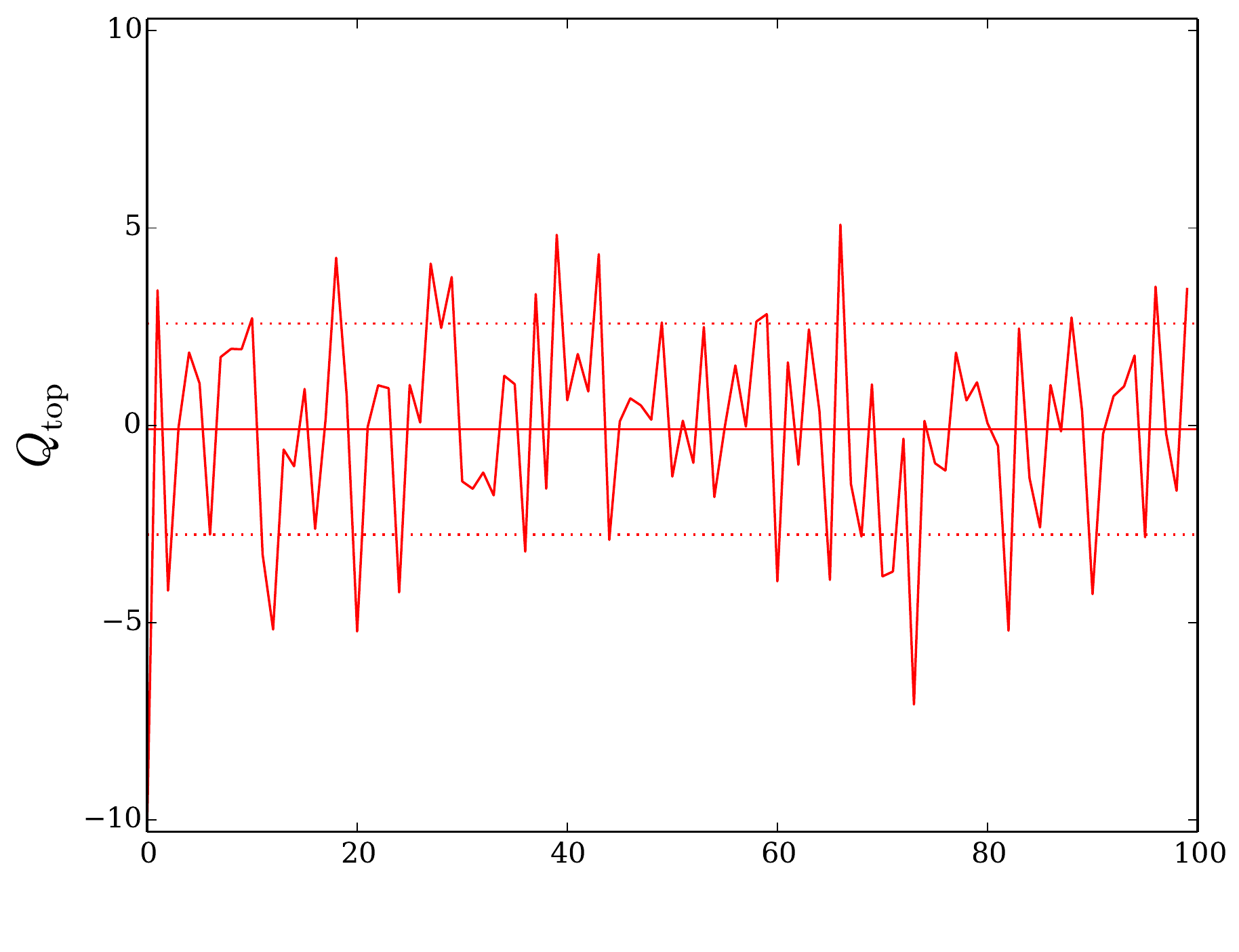} \includegraphics[width=.32\textwidth]{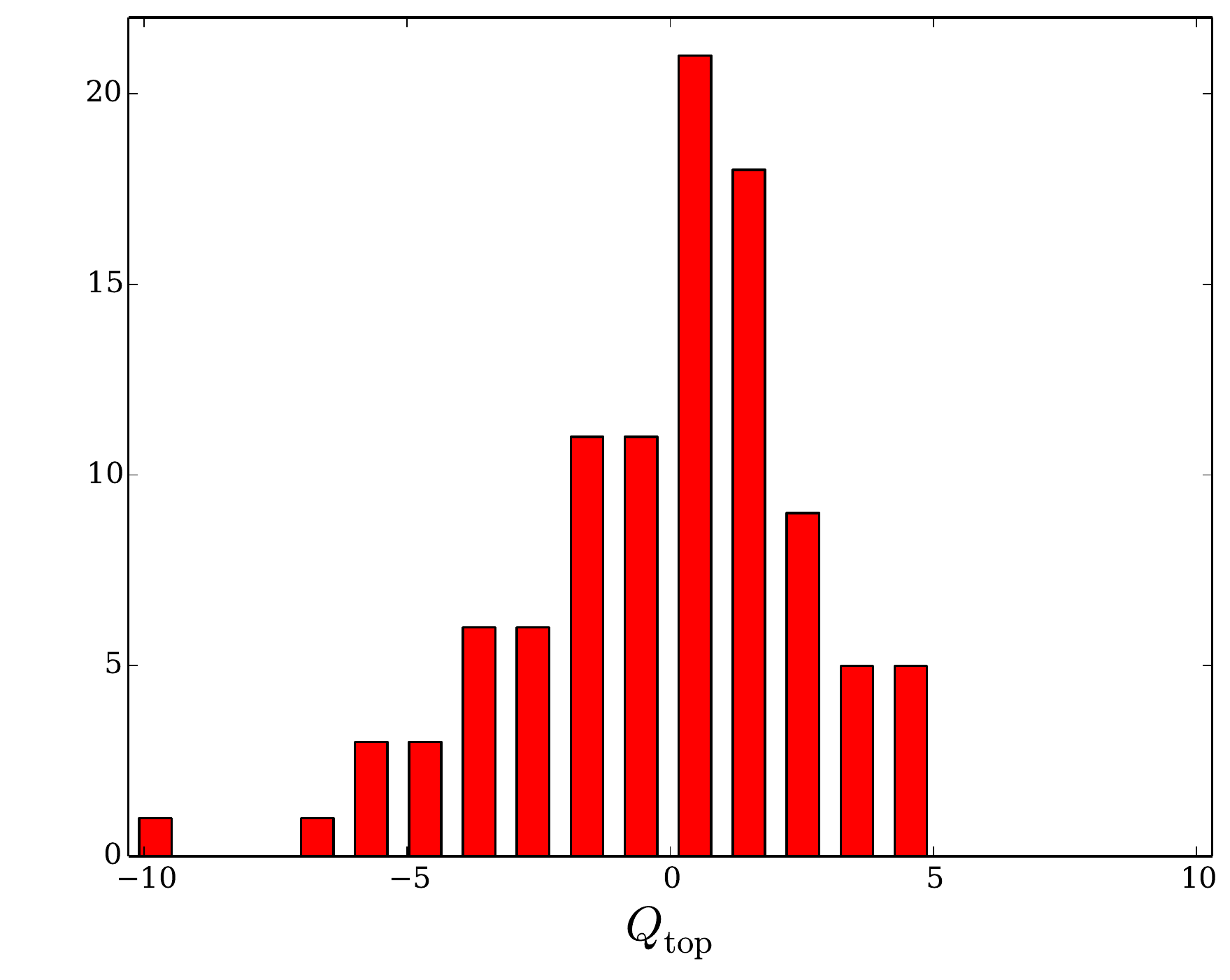}\\
    \includegraphics[width=.32\textwidth]{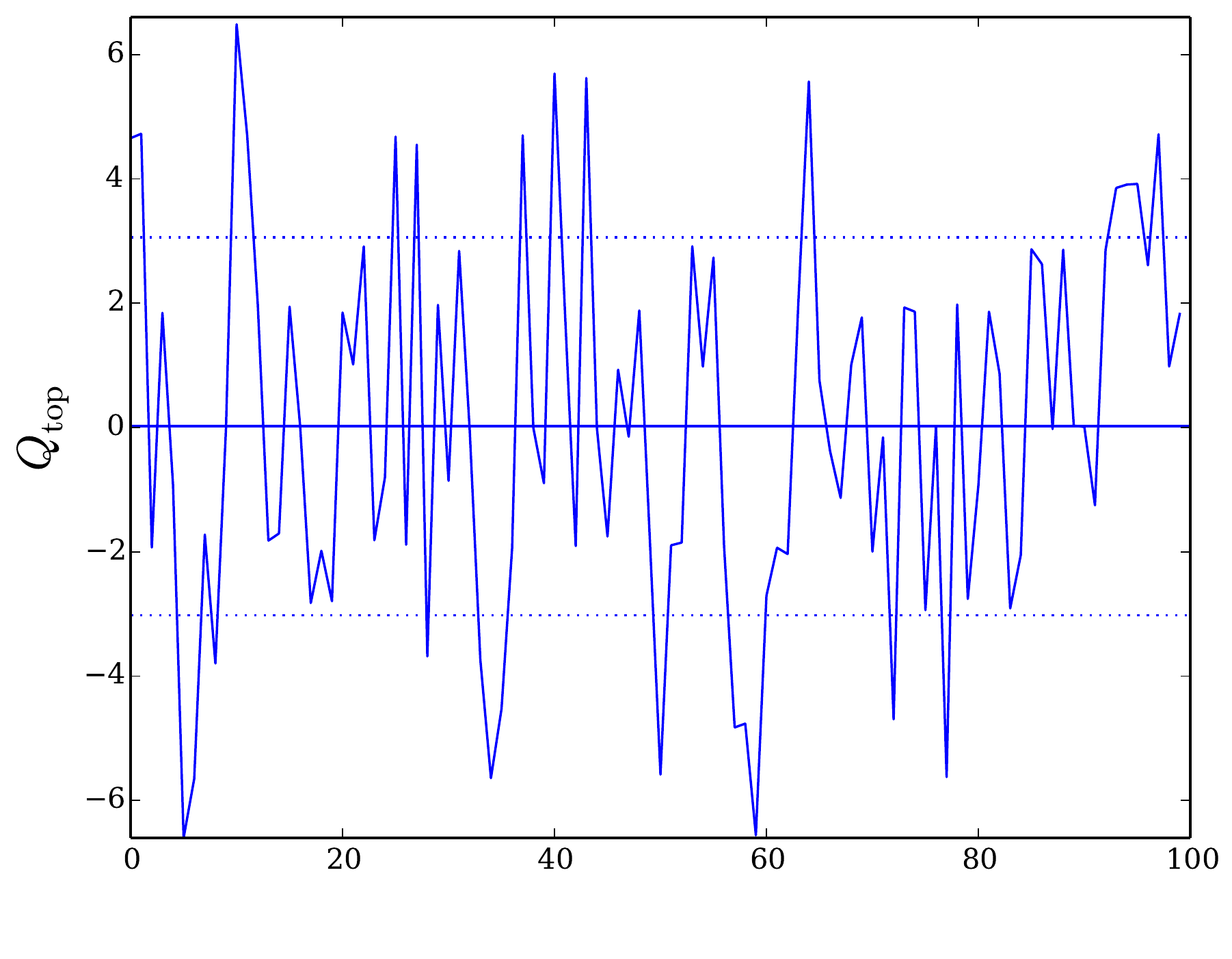} \includegraphics[width=.32\textwidth]{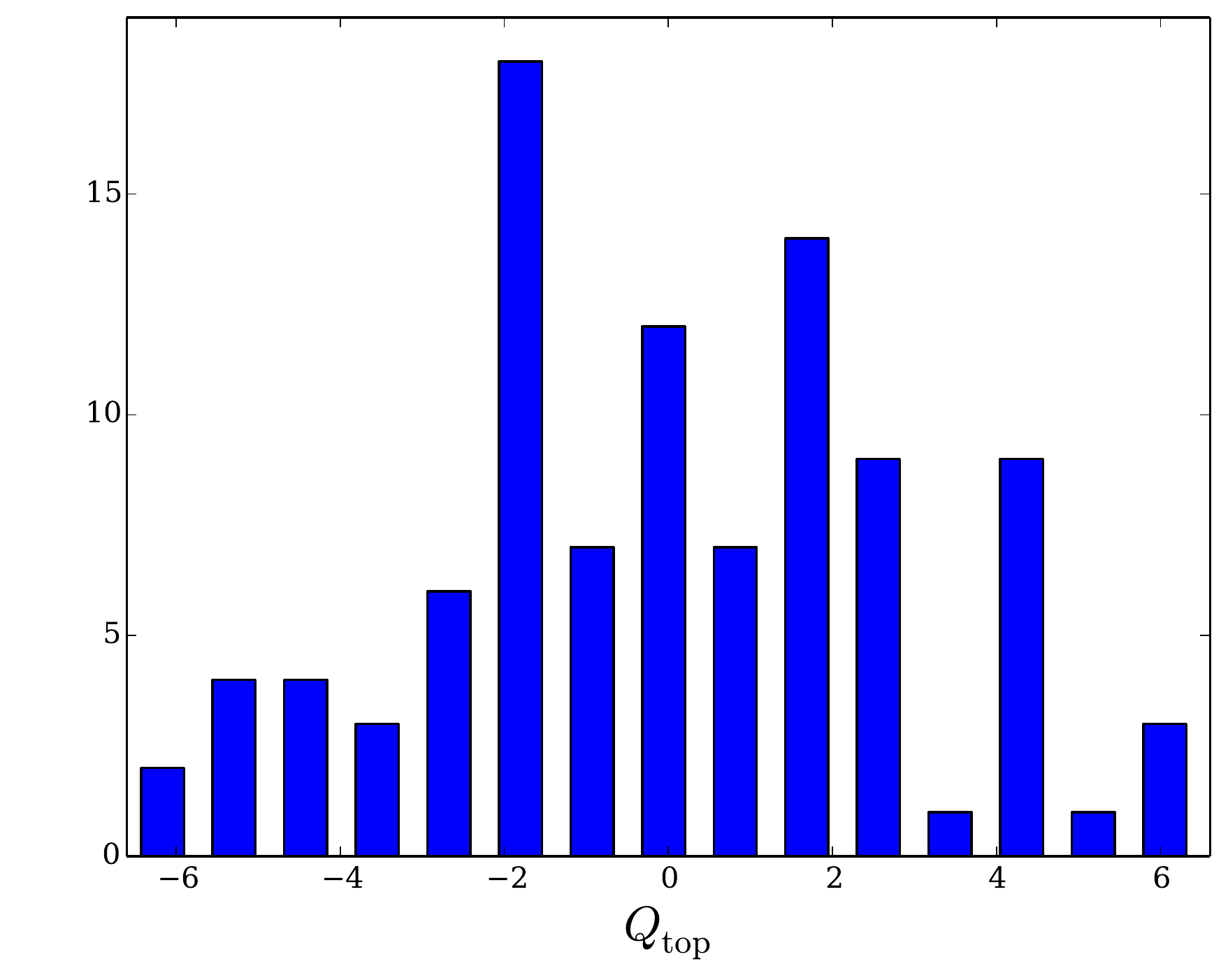}\\
    \includegraphics[width=.32\textwidth]{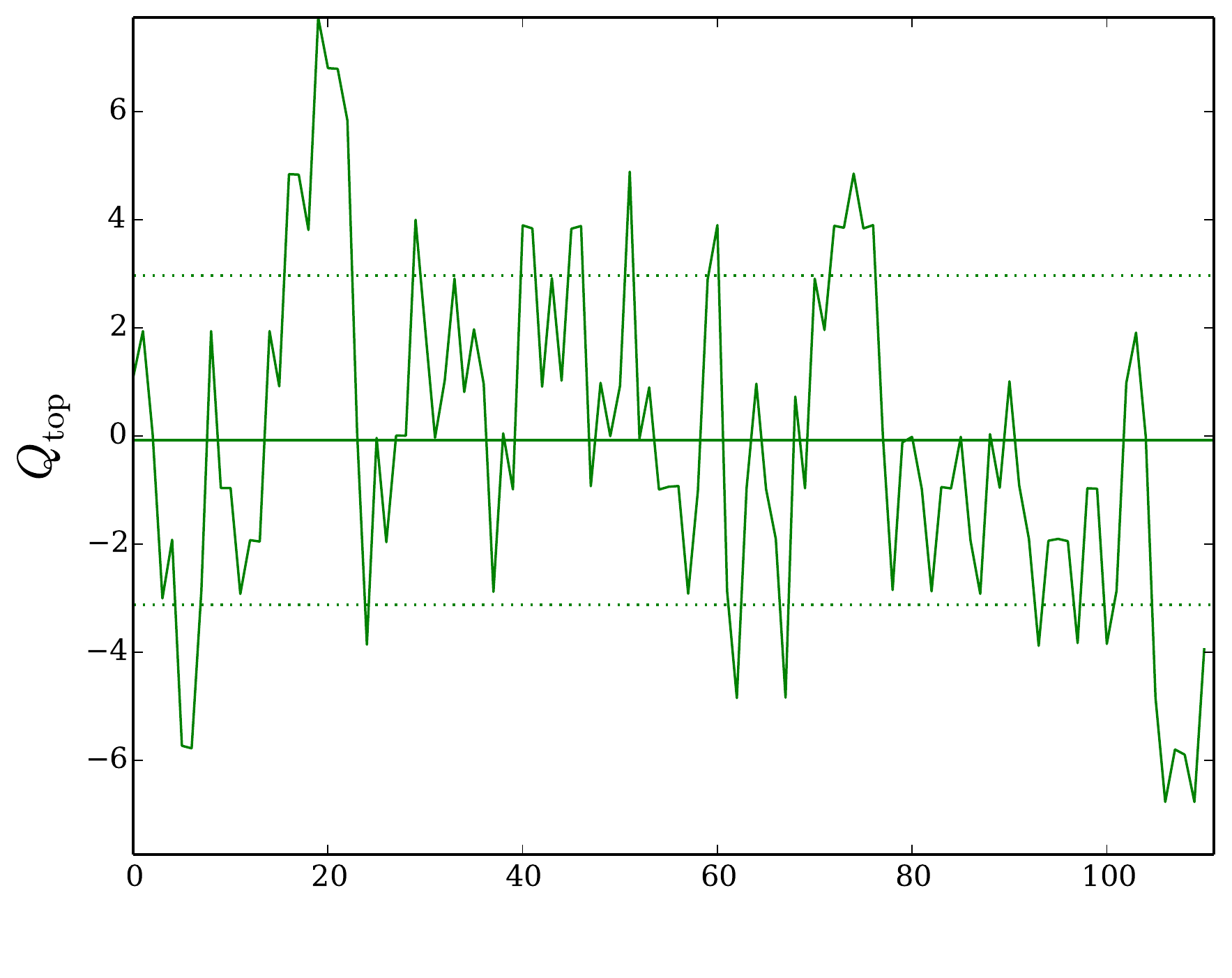} \includegraphics[width=.32\textwidth]{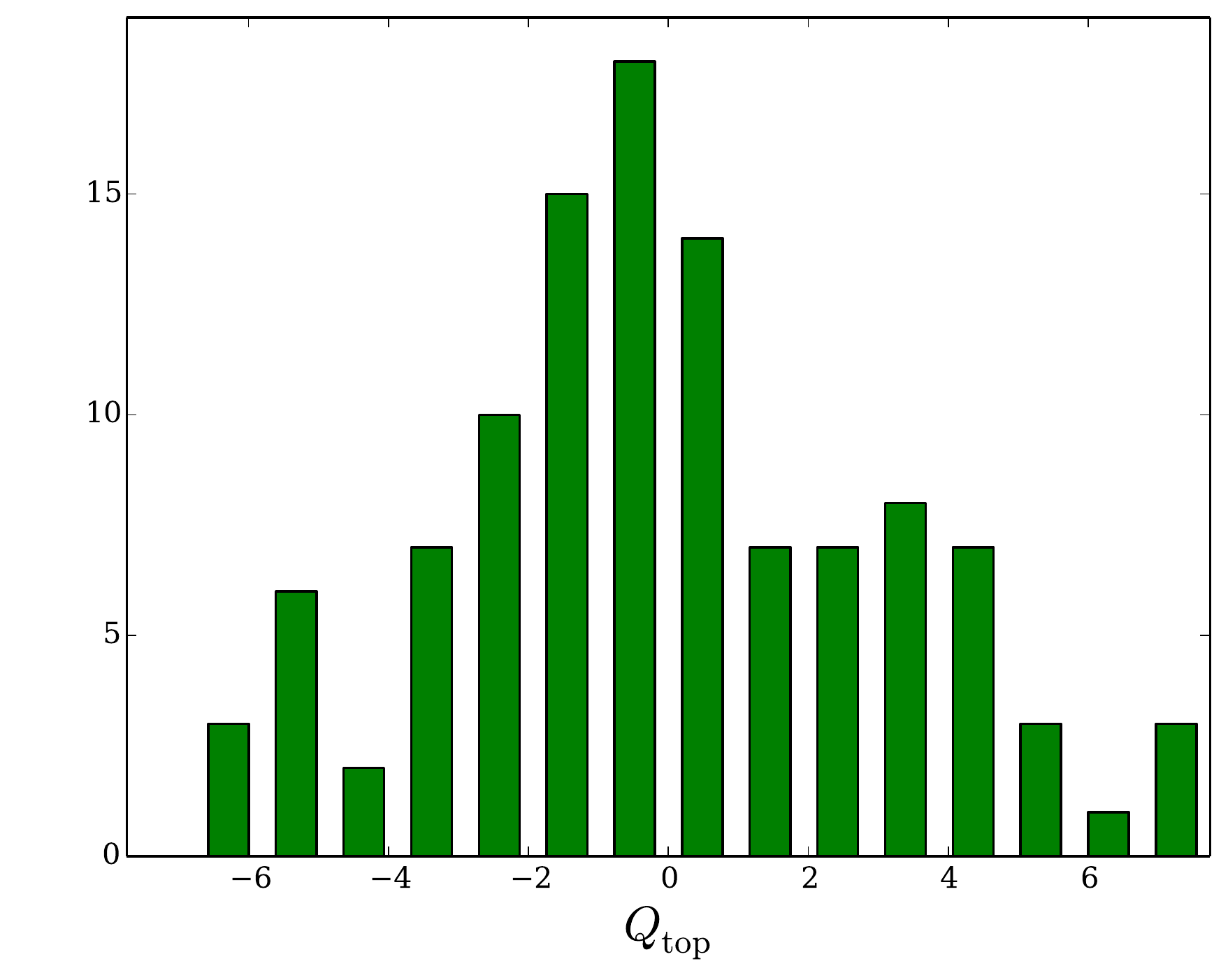}\\
    \includegraphics[width=.32\textwidth]{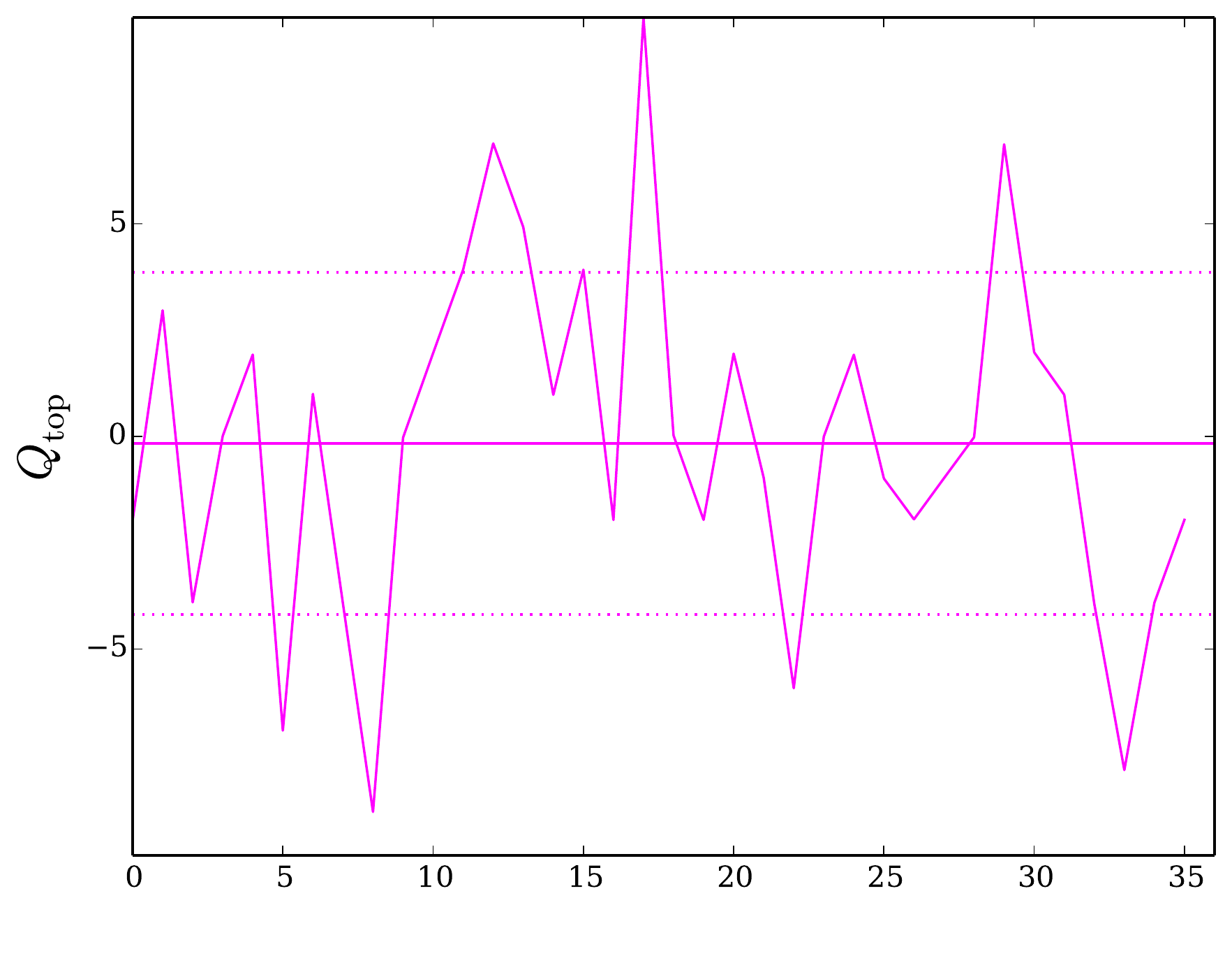} \includegraphics[width=.32\textwidth]{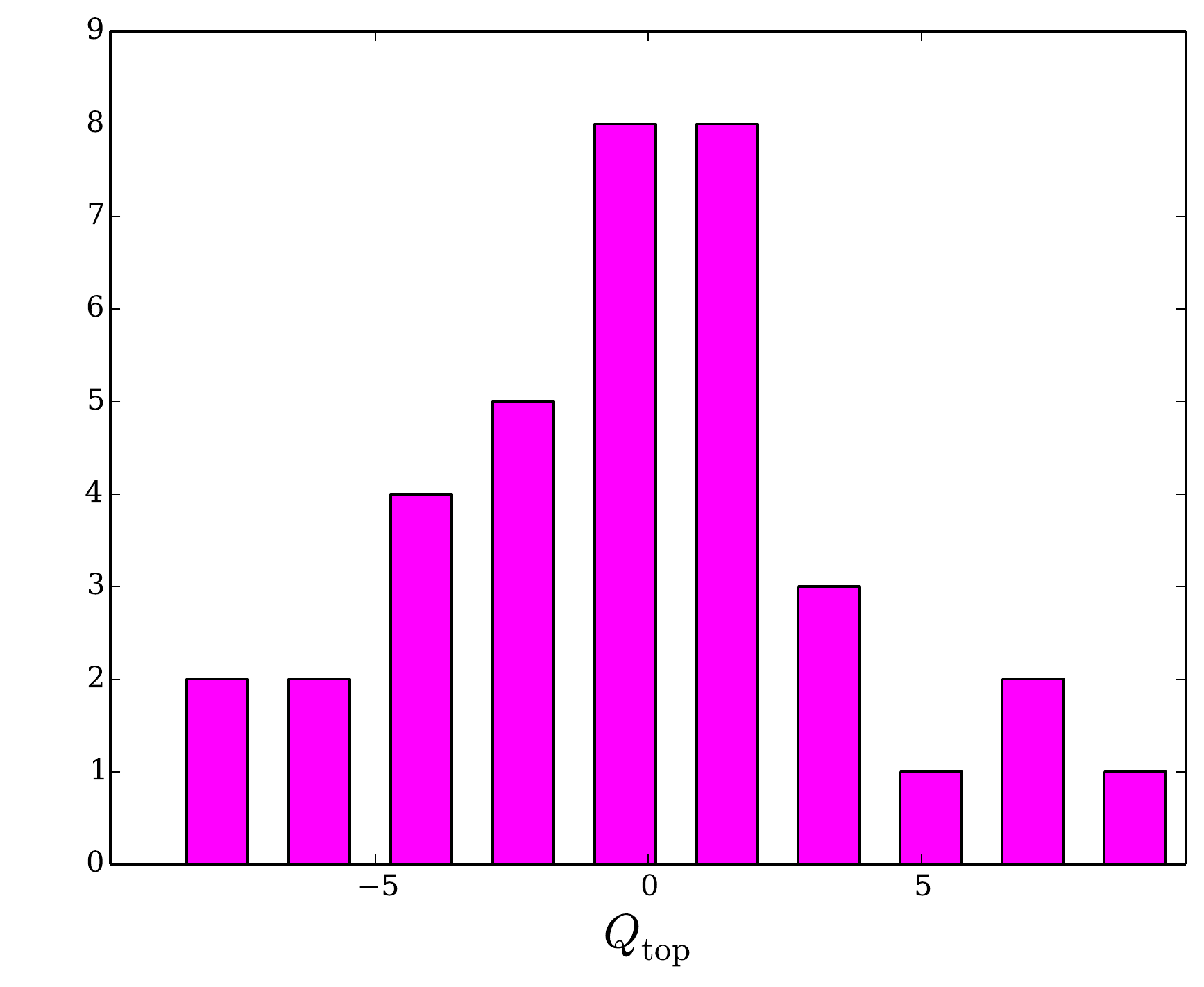}\\
    \caption{Topological charge evolution (left) and histograms
      (right) for the ensembles listed in table~\ref{tab:runpars},
      \ref{tab:lat_vol} and~\ref{tab:autocorr}. This plot only
      includes configurations that are decorrelated.  The lattice
      spacing decreases from top to bottom.\label{fig:topoloical
        charge evolution}}
\end{figure}
\clearpage

\section{Correlator fit results}
\subsection{Decay constant data}
\begin{table} 
 \center \small 
\label{tab:dec-l16t32} 
 \begin{tabular}{|l|c c c |c c c |} 
  \hline tag & $am$ & $af$ & $\chi^2/\mathrm{d.o.f.}$ & $am$ & $af$ & $\chi^2/\mathrm{d.o.f.}$ \\ & \multicolumn{3}{c|}{$am_s=0.036$} & \multicolumn{3}{c|}{$am_s=0.037$} \\ 
   \hline 
   ss & \color{black}{0.3433(16)} & \color{black}{0.11309(57)} & \color{black}{0.6033} & \color{black}{0.3479(16)} & \color{black}{0.11360(57)} & \color{black}{0.5384} \\ 
   sh0 & \color{black}{0.4738(15)} & \color{black}{0.12710(81)} & \color{black}{0.3443} & \color{black}{0.4764(12)} & \color{black}{0.12775(73)} & \color{black}{0.4052} \\ 
   sh1 & \color{black}{0.5628(14)} & \color{black}{0.13452(83)} & \color{black}{0.3609} & \color{black}{0.5654(12)} & \color{black}{0.13519(80)} & \color{black}{0.4885} \\ 
   sh2 & \color{black}{0.6448(14)} & \color{black}{0.13960(85)} & \color{black}{0.3471} & \color{black}{0.6463(13)} & \color{black}{0.13991(85)} & \color{black}{0.3455} \\ 
   sh3 & \color{black}{0.7215(13)} & \color{black}{0.14292(88)} & \color{black}{0.3456} & \color{black}{0.7227(14)} & \color{black}{0.14314(90)} & \color{black}{0.3522} \\ 
   sh4 & \color{black}{0.7938(14)} & \color{black}{0.14480(92)} & \color{black}{0.3825} & \color{black}{0.7953(13)} & \color{black}{0.14501(91)} & \color{black}{0.2996} \\ 
   sh5 & \color{black}{0.8614(15)} & \color{black}{0.1447(10)} & \color{black}{0.2965} & \color{black}{0.8628(15)} & \color{black}{0.1451(10)} & \color{black}{0.2925} \\ 
   sh6 & \color{black}{0.9254(16)} & \color{black}{0.1439(11)} & \color{black}{0.3610} & \color{black}{0.9266(15)} & \color{black}{0.1443(10)} & \color{black}{0.3577} \\ 
    \hline \end{tabular} \caption{Fit results for strange-strange and strange-heavy pseudoscalar masses and decay constants in lattice units for the ensembles $\beta=4.41$.} 
 
\end{table}
\begin{table} 
 \center \small 
\label{tab:dec-l24t48} 
 \begin{tabular}{|l|c c c |c c c |} 
  \hline tag & $am$ & $af$ & $\chi^2/\mathrm{d.o.f.}$ & $am$ & $af$ & $\chi^2/\mathrm{d.o.f.}$ \\ & \multicolumn{3}{c|}{$am_s=0.024$} & \multicolumn{3}{c|}{$am_s=0.026$} \\ 
   \hline 
   ss & \color{black}{0.23894(93)} & \color{black}{0.07812(48)} & \color{black}{0.7714} & \color{black}{0.24883(90)} & \color{black}{0.07926(47)} & \color{black}{0.8466} \\ 
   sh0 & \color{black}{0.3294(11)} & \color{black}{0.08816(86)} & \color{black}{0.8914} & \color{black}{0.3333(11)} & \color{black}{0.08871(85)} & \color{black}{0.9314} \\ 
   sh1 & \color{black}{0.3919(11)} & \color{black}{0.09311(83)} & \color{black}{0.9179} & \color{black}{0.3955(11)} & \color{black}{0.09381(84)} & \color{black}{0.9530} \\ 
   sh2 & \color{black}{0.4494(10)} & \color{black}{0.09678(84)} & \color{black}{0.9083} & \color{black}{0.4528(11)} & \color{black}{0.09766(90)} & \color{black}{0.9559} \\ 
   sh3 & \color{black}{0.5033(10)} & \color{black}{0.09930(84)} & \color{black}{0.9123} & \color{black}{0.5063(11)} & \color{black}{0.09997(86)} & \color{black}{0.9319} \\ 
   sh4 & \color{black}{0.5545(10)} & \color{black}{0.10099(84)} & \color{black}{0.8906} & \color{black}{0.5573(10)} & \color{black}{0.10167(82)} & \color{black}{0.8719} \\ 
   sh5 & \color{black}{0.6033(10)} & \color{black}{0.10200(84)} & \color{black}{0.8382} & \color{black}{0.6059(10)} & \color{black}{0.10269(82)} & \color{black}{0.8245} \\ 
   sh6 & \color{black}{0.6491(13)} & \color{black}{0.1019(10)} & \color{black}{0.8190} & \color{black}{0.6518(12)} & \color{black}{0.10272(97)} & \color{black}{0.8216} \\ 
   sh7 & \color{black}{0.6931(13)} & \color{black}{0.10114(92)} & \color{black}{0.8668} & \color{black}{0.6962(13)} & \color{black}{0.10248(98)} & \color{black}{0.7454} \\ 
   sh8 & \color{black}{0.7357(13)} & \color{black}{0.10037(93)} & \color{black}{0.7866} & \color{black}{0.7387(13)} & \color{black}{0.1016(10)} & \color{black}{0.6989} \\ 
   sh9 & \color{black}{0.7768(15)} & \color{black}{0.0996(11)} & \color{black}{0.6151} & \color{black}{0.7793(14)} & \color{black}{0.1004(10)} & \color{black}{0.6358} \\ 
   sh10 & \color{black}{0.8151(15)} & \color{black}{0.0973(10)} & \color{black}{0.6496} & \color{black}{0.8181(15)} & \color{black}{0.0984(11)} & \color{black}{0.5990} \\ 
    \hline \end{tabular} \caption{Fit results for strange-strange and strange-heavy pseudoscalar masses and decay constants in lattice units for the ensembles $\beta=4.66$.} 
 
\end{table}
\begin{table} 
 \center \small 
\label{tab:dec-l32t64} 
 \begin{tabular}{|l|c c c |c c c |} 
  \hline tag & $am$ & $af$ & $\chi^2/\mathrm{d.o.f.}$ & $am$ & $af$ & $\chi^2/\mathrm{d.o.f.}$ \\ & \multicolumn{3}{c|}{$am_s=0.018$} & \multicolumn{3}{c|}{$am_s=0.020$} \\ 
   \hline 
   ss & \color{black}{0.1773(10)} & \color{black}{0.05630(37)} & \color{black}{0.9652} & \color{black}{0.1872(10)} & \color{black}{0.05739(36)} & \color{black}{0.8819} \\ 
   sh0 & \color{black}{0.28568(95)} & \color{black}{0.06578(66)} & \color{black}{0.7877} & \color{black}{0.28925(85)} & \color{black}{0.06647(60)} & \color{black}{0.6169} \\ 
   sh1 & \color{black}{0.35566(88)} & \color{black}{0.06991(68)} & \color{black}{0.5133} & \color{black}{0.35889(79)} & \color{black}{0.07058(60)} & \color{black}{0.4933} \\ 
   sh2 & \color{black}{0.41960(90)} & \color{black}{0.07189(70)} & \color{black}{0.4931} & \color{black}{0.42283(84)} & \color{black}{0.07277(66)} & \color{black}{0.4966} \\ 
   sh3 & \color{black}{0.47922(98)} & \color{black}{0.07256(76)} & \color{black}{0.4981} & \color{black}{0.48239(90)} & \color{black}{0.07352(70)} & \color{black}{0.5273} \\ 
   sh4 & \color{black}{0.53492(93)} & \color{black}{0.07213(78)} & \color{black}{0.7555} & \color{black}{0.53775(89)} & \color{black}{0.07311(74)} & \color{black}{0.7476} \\ 
   sh5 & \color{black}{0.58831(97)} & \color{black}{0.07138(80)} & \color{black}{0.7290} & \color{black}{0.59103(92)} & \color{black}{0.07241(76)} & \color{black}{0.7275} \\ 
   sh6 & \color{black}{0.6391(10)} & \color{black}{0.07023(81)} & \color{black}{0.7054} & \color{black}{0.6419(10)} & \color{black}{0.07130(80)} & \color{black}{0.7394} \\ 
   sh7 & \color{black}{0.6872(10)} & \color{black}{0.06876(83)} & \color{black}{0.6944} & \color{black}{0.6899(10)} & \color{black}{0.06962(82)} & \color{black}{0.7975} \\ 
   sh8 & \color{black}{0.73280(96)} & \color{black}{0.06700(65)} & \color{black}{0.7404} & \color{black}{0.7352(10)} & \color{black}{0.06809(85)} & \color{black}{0.6866} \\ 
    \hline \end{tabular} \caption{Fit results for strange-strange and strange-heavy pseudoscalar masses and decay constants in lattice units for the ensembles $\beta=4.89$.} 
 
\end{table}
\begin{table} 
 \center \small 
\label{tab:dec-l48t96} 
 \begin{tabular}{|l|c c c |c c c |} 
  \hline tag & $am$ & $af$ & $\chi^2/\mathrm{d.o.f.}$ & $am$ & $af$ & $\chi^2/\mathrm{d.o.f.}$ \\ & \multicolumn{3}{c|}{$am_s=0.0118$} & \multicolumn{3}{c|}{$am_s=0.0133$} \\ 
   \hline 
   ss & \color{black}{0.1212(12)} & \color{black}{0.03723(73)} & \color{black}{0.0189} & \color{black}{0.1288(12)} & \color{black}{0.0382(14)} & \color{black}{0.0080} \\ 
   sh0 & \color{black}{0.1815(13)} & \color{black}{0.04233(96)} & \color{black}{0.0120} & \color{black}{0.1841(13)} & \color{black}{0.04265(98)} & \color{black}{0.0005} \\ 
   sh1 & \color{black}{0.2528(12)} & \color{black}{0.04615(91)} & \color{black}{0.0033} & \color{black}{0.2552(11)} & \color{black}{0.04659(87)} & \color{black}{0.0042} \\ 
   sh2 & \color{black}{0.3167(12)} & \color{black}{0.04795(88)} & \color{black}{0.0051} & \color{black}{0.3189(12)} & \color{black}{0.04843(84)} & \color{black}{0.0061} \\ 
   sh3 & \color{black}{0.3758(15)} & \color{black}{0.0485(10)} & \color{black}{0.0028} & \color{black}{0.3778(14)} & \color{black}{0.04891(99)} & \color{black}{0.0027} \\ 
   sh4 & \color{black}{0.4315(16)} & \color{black}{0.0484(10)} & \color{black}{0.0034} & \color{black}{0.4335(14)} & \color{black}{0.0488(10)} & \color{black}{0.0032} \\ 
   sh5 & \color{black}{0.4843(17)} & \color{black}{0.0478(10)} & \color{black}{0.0041} & \color{black}{0.4862(15)} & \color{black}{0.0483(10)} & \color{black}{0.0038} \\ 
   sh6 & \color{black}{0.5344(18)} & \color{black}{0.0470(10)} & \color{black}{0.0052} & \color{black}{0.5362(16)} & \color{black}{0.0474(10)} & \color{black}{0.0051} \\ 
    \hline \end{tabular} \caption{Fit results for strange-strange and strange-heavy pseudoscalar masses and decay constants in lattice units for the ensembles $\beta=5.20$.} 
 
\end{table}

\clearpage
\subsection{Dispersion relation data}
\begin{table} 
 \center \small 
\label{tab:disp-l16t32} 
 \begin{tabular}{|l|c c c c |c c c c |} 
  \hline $\mathbf{n}$ & $(0,0,0)$ & $(1,0,0)$ & $(1,1,0)$ & $(1,1,1)$ & $(0,0,0)$ & $(1,0,0)$ & $(1,1,0)$ & $(1,1,1)$ \\ & \multicolumn{4}{c|}{$am_s=0.034$} & \multicolumn{4}{c|}{$am_s=0.036$} \\ 
   \hline 
   ss & \color{black}{0.3321(19)} & - & - & - & \color{black}{0.3415(18)} & - & - & - \\ 
   sh0 & \color{black}{0.4699(15)} & \color{black}{0.619(12)} & \color{black}{0.755(16)} & \color{black}{0.725(68)} & \color{black}{0.4735(14)} & \color{black}{0.621(12)} & \color{black}{0.757(15)} & \color{black}{0.729(65)} \\ 
   sh1 & \color{black}{0.5591(14)} & \color{black}{0.686(10)} & \color{black}{0.800(14)} & \color{black}{0.793(39)} & \color{black}{0.5624(14)} & \color{black}{0.689(10)} & \color{black}{0.801(13)} & \color{black}{0.796(38)} \\ 
   sh2 & \color{black}{0.6410(14)} & \color{black}{0.750(10)} & \color{black}{0.852(12)} & \color{black}{0.853(32)} & \color{black}{0.6440(14)} & \color{black}{0.7528(98)} & \color{black}{0.854(11)} & \color{black}{0.856(31)} \\ 
   sh3 & \color{black}{0.7174(14)} & \color{black}{0.8143(89)} & \color{black}{0.904(10)} & \color{black}{0.917(25)} & \color{black}{0.7203(14)} & \color{black}{0.8166(87)} & \color{black}{0.906(10)} & \color{black}{0.920(24)} \\ 
   sh4 & \color{black}{0.7894(14)} & \color{black}{0.8763(82)} & \color{black}{0.9566(94)} & \color{black}{0.977(21)} & \color{black}{0.7922(13)} & \color{black}{0.8787(80)} & \color{black}{0.9588(92)} & \color{black}{0.980(20)} \\ 
   sh5 & \color{black}{0.8572(14)} & \color{black}{0.9367(78)} & \color{black}{1.0049(90)} & \color{black}{1.033(18)} & \color{black}{0.8599(14)} & \color{black}{0.9389(77)} & \color{black}{1.0072(87)} & \color{black}{1.035(18)} \\ 
   sh6 & \color{black}{0.9209(14)} & \color{black}{0.9935(76)} & \color{black}{1.0561(84)} & \color{black}{1.086(17)} & \color{black}{0.9235(14)} & \color{black}{0.9957(75)} & \color{black}{1.0584(81)} & \color{black}{1.088(16)} \\ 
    \hline \end{tabular} \caption{Fit results for the energy of strange-strange and strange-heavy pseudoscalar mesons in lattice units as a function of the momentum for the ensemble $\beta=4.41$.} 
 
\end{table}
\begin{table} 
 \center \small 
\label{tab:disp-l24t48} 
 \begin{tabular}{|l|c c c c |c c c c |} 
  \hline $\mathbf{n}$ & $(0,0,0)$ & $(1,0,0)$ & $(1,1,0)$ & $(1,1,1)$ & $(0,0,0)$ & $(1,0,0)$ & $(1,1,0)$ & $(1,1,1)$ \\ & \multicolumn{4}{c|}{$am_s=0.024$} & \multicolumn{4}{c|}{$am_s=0.026$} \\ 
   \hline 
   ss & \color{black}{0.23749(89)} & - & - & - & \color{black}{0.24723(86)} & - & - & - \\ 
   sh0 & \color{black}{0.32942(73)} & \color{black}{0.419(16)} & \color{black}{0.477(19)} & \color{black}{0.583(26)} & \color{black}{0.33328(71)} & \color{black}{0.422(15)} & \color{black}{0.480(19)} & \color{black}{0.585(25)} \\ 
   sh1 & \color{black}{0.39143(72)} & \color{black}{0.473(12)} & \color{black}{0.529(14)} & \color{black}{0.615(32)} & \color{black}{0.39492(71)} & \color{black}{0.476(12)} & \color{black}{0.532(13)} & \color{black}{0.617(31)} \\ 
   sh2 & \color{black}{0.44837(73)} & \color{black}{0.524(11)} & \color{black}{0.577(13)} & \color{black}{0.650(42)} & \color{black}{0.45163(72)} & \color{black}{0.527(10)} & \color{black}{0.579(12)} & \color{black}{0.652(40)} \\ 
   sh3 & \color{black}{0.50185(74)} & \color{black}{0.572(10)} & \color{black}{0.620(10)} & \color{black}{0.668(48)} & \color{black}{0.50496(72)} & \color{black}{0.5744(98)} & \color{black}{0.621(10)} & \color{black}{0.672(46)} \\ 
   sh4 & \color{black}{0.55252(76)} & \color{black}{0.6178(96)} & \color{black}{0.6603(94)} & \color{black}{0.699(40)} & \color{black}{0.55552(74)} & \color{black}{0.6206(93)} & \color{black}{0.6629(89)} & \color{black}{0.702(38)} \\ 
   sh5 & \color{black}{0.60091(78)} & \color{black}{0.6623(91)} & \color{black}{0.7010(85)} & \color{black}{0.724(39)} & \color{black}{0.60382(76)} & \color{black}{0.6651(88)} & \color{black}{0.7036(81)} & \color{black}{0.727(37)} \\ 
   sh6 & \color{black}{0.64723(81)} & \color{black}{0.7054(87)} & \color{black}{0.7407(78)} & \color{black}{0.757(33)} & \color{black}{0.65007(78)} & \color{black}{0.7081(84)} & \color{black}{0.7433(74)} & \color{black}{0.761(32)} \\ 
   sh7 & \color{black}{0.69161(83)} & \color{black}{0.7468(84)} & \color{black}{0.7792(73)} & \color{black}{0.790(29)} & \color{black}{0.69440(80)} & \color{black}{0.7495(82)} & \color{black}{0.7819(69)} & \color{black}{0.794(28)} \\ 
   sh8 & \color{black}{0.73410(87)} & \color{black}{0.7868(82)} & \color{black}{0.8166(69)} & \color{black}{0.823(26)} & \color{black}{0.73685(84)} & \color{black}{0.7895(80)} & \color{black}{0.8192(65)} & \color{black}{0.827(25)} \\ 
   sh9 & \color{black}{0.77473(90)} & \color{black}{0.8264(84)} & \color{black}{0.8526(66)} & \color{black}{0.856(24)} & \color{black}{0.77744(86)} & \color{black}{0.8291(81)} & \color{black}{0.8553(62)} & \color{black}{0.859(23)} \\ 
   sh10 & \color{black}{0.81346(93)} & \color{black}{0.8632(83)} & \color{black}{0.8873(63)} & \color{black}{0.881(24)} & \color{black}{0.81614(89)} & \color{black}{0.8659(80)} & \color{black}{0.8899(60)} & \color{black}{0.885(23)} \\ 
    \hline \end{tabular} \caption{Fit results for the energy of strange-strange and strange-heavy pseudoscalar mesons in lattice units as a function of the momentum for the ensemble $\beta=4.66$.} 
 
\end{table}
\begin{table} 
 \center \small 
\label{tab:disp-l32t64} 
 \begin{tabular}{|l|c c c c |c c c c |} 
  \hline $\mathbf{n}$ & $(0,0,0)$ & $(1,0,0)$ & $(1,1,0)$ & $(1,1,1)$ & $(0,0,0)$ & $(1,0,0)$ & $(1,1,0)$ & $(1,1,1)$ \\ & \multicolumn{4}{c|}{$am_s=0.018$} & \multicolumn{4}{c|}{$am_s=0.020$} \\ 
   \hline 
   ss & \color{black}{0.17758(75)} & - & - & - & \color{black}{0.18735(73)} & - & - & - \\ 
   sh0 & \color{black}{0.28700(56)} & \color{black}{0.350(10)} & \color{black}{0.4032(50)} & \color{black}{0.430(23)} & \color{black}{0.29061(54)} & \color{black}{0.3529(97)} & \color{black}{0.4058(48)} & \color{black}{0.432(22)} \\ 
   sh1 & \color{black}{0.35690(58)} & \color{black}{0.4085(78)} & \color{black}{0.4562(36)} & \color{black}{0.480(16)} & \color{black}{0.36016(56)} & \color{black}{0.4110(74)} & \color{black}{0.4588(34)} & \color{black}{0.483(15)} \\ 
   sh2 & \color{black}{0.42074(60)} & \color{black}{0.4653(66)} & \color{black}{0.5063(35)} & \color{black}{0.530(13)} & \color{black}{0.42380(58)} & \color{black}{0.4678(63)} & \color{black}{0.5089(33)} & \color{black}{0.533(12)} \\ 
   sh3 & \color{black}{0.48047(61)} & \color{black}{0.5186(63)} & \color{black}{0.5542(36)} & \color{black}{0.579(11)} & \color{black}{0.48339(58)} & \color{black}{0.5211(60)} & \color{black}{0.5567(34)} & \color{black}{0.582(11)} \\ 
   sh4 & \color{black}{0.53680(65)} & \color{black}{0.5717(59)} & \color{black}{0.6029(32)} & \color{black}{0.6257(90)} & \color{black}{0.53964(62)} & \color{black}{0.5742(56)} & \color{black}{0.6054(31)} & \color{black}{0.6285(87)} \\ 
   sh5 & \color{black}{0.59020(67)} & \color{black}{0.6222(57)} & \color{black}{0.6502(30)} & \color{black}{0.6727(84)} & \color{black}{0.59298(64)} & \color{black}{0.6247(54)} & \color{black}{0.6527(28)} & \color{black}{0.6755(81)} \\ 
   sh6 & \color{black}{0.64086(69)} & \color{black}{0.6697(58)} & \color{black}{0.6958(28)} & \color{black}{0.7180(80)} & \color{black}{0.64359(66)} & \color{black}{0.6722(55)} & \color{black}{0.6983(26)} & \color{black}{0.7208(77)} \\ 
   sh7 & \color{black}{0.68882(72)} & \color{black}{0.7161(55)} & \color{black}{0.7398(26)} & \color{black}{0.7620(84)} & \color{black}{0.69150(68)} & \color{black}{0.7186(53)} & \color{black}{0.7423(24)} & \color{black}{0.7648(80)} \\ 
   sh8 & \color{black}{0.73401(74)} & \color{black}{0.7598(56)} & \color{black}{0.7814(25)} & \color{black}{0.8031(83)} & \color{black}{0.73667(70)} & \color{black}{0.7623(53)} & \color{black}{0.7840(23)} & \color{black}{0.8059(79)} \\ 
    \hline \end{tabular} \caption{Fit results for the energy of strange-strange and strange-heavy pseudoscalar mesons in lattice units as a function of the momentum for the ensemble $\beta=4.89$.} 
 
\end{table}
\begin{table} 
 \center \small 
\label{tab:disp-l48t96} 
 \begin{tabular}{|l|c c c c |c c c c |} 
  \hline $\mathbf{n}$ & $(0,0,0)$ & $(1,0,0)$ & $(1,1,0)$ & $(1,1,1)$ & $(0,0,0)$ & $(1,0,0)$ & $(1,1,0)$ & $(1,1,1)$ \\ & \multicolumn{4}{c|}{$am_s=0.011$} & \multicolumn{4}{c|}{$am_s=0.013$} \\ 
   \hline 
   ss & \color{black}{0.1143(13)} & - & - & - & \color{black}{0.1246(11)} & - & - & - \\ 
   sh0 & \color{black}{0.17940(63)} & \color{black}{0.246(13)} & \color{black}{0.2454(78)} & \color{black}{0.269(16)} & \color{black}{0.18322(60)} & \color{black}{0.248(12)} & \color{black}{0.2495(74)} & \color{black}{0.272(15)} \\ 
   sh1 & \color{black}{0.25110(52)} & \color{black}{0.302(12)} & \color{black}{0.3035(51)} & \color{black}{0.327(14)} & \color{black}{0.25441(49)} & \color{black}{0.304(11)} & \color{black}{0.3074(47)} & \color{black}{0.330(13)} \\ 
   sh2 & \color{black}{0.31470(52)} & \color{black}{0.358(11)} & \color{black}{0.3590(43)} & \color{black}{0.374(13)} & \color{black}{0.31778(49)} & \color{black}{0.360(10)} & \color{black}{0.3625(40)} & \color{black}{0.377(12)} \\ 
   sh3 & \color{black}{0.37356(56)} & \color{black}{0.412(11)} & \color{black}{0.4116(38)} & \color{black}{0.430(13)} & \color{black}{0.37651(52)} & \color{black}{0.414(10)} & \color{black}{0.4148(34)} & \color{black}{0.433(12)} \\ 
   sh4 & \color{black}{0.42894(61)} & \color{black}{0.465(11)} & \color{black}{0.4625(36)} & \color{black}{0.477(11)} & \color{black}{0.43180(58)} & \color{black}{0.466(10)} & \color{black}{0.4656(32)} & \color{black}{0.481(11)} \\ 
   sh5 & \color{black}{0.48138(67)} & \color{black}{0.516(10)} & \color{black}{0.5114(35)} & \color{black}{0.524(10)} & \color{black}{0.48419(62)} & \color{black}{0.517(10)} & \color{black}{0.5143(31)} & \color{black}{0.527(10)} \\ 
   sh6 & \color{black}{0.53114(72)} & \color{black}{0.564(11)} & \color{black}{0.5582(34)} & \color{black}{0.5689(99)} & \color{black}{0.53390(67)} & \color{black}{0.566(10)} & \color{black}{0.5611(30)} & \color{black}{0.5723(95)} \\ 
    \hline \end{tabular} \caption{Fit results for the energy of strange-strange and strange-heavy pseudoscalar mesons in lattice units as a function of the momentum for the ensemble $\beta=5.20$.} 
 
\end{table}

\clearpage
\bibliographystyle{JHEP}
\bibliography{biblio}

\providecommand{\href}[2]{#2}\begingroup\raggedright\begin{thebibliography}{10}

\bibitem{Lepage:1987gg}
G.~P. Lepage and B.~A. Thacker, \emph{{Effective Lagrangians for Simulating
  Heavy Quark Systems}},
  \href{http://dx.doi.org/10.1016/0920-5632(88)90102-8}{\emph{Nucl. Phys. Proc.
  Suppl.} {\bf 4} (1988) 199}.

\bibitem{Thacker:1990bm}
B.~A. Thacker and G.~P. Lepage, \emph{{Heavy quark bound states in lattice
  QCD}}, \href{http://dx.doi.org/10.1103/PhysRevD.43.196}{\emph{Phys. Rev.}
  {\bf D43} (1991) 196--208}.

\bibitem{Eichten:1989zv}
E.~Eichten and B.~R. Hill, \emph{{An Effective Field Theory for the Calculation
  of Matrix Elements Involving Heavy Quarks}},
  \href{http://dx.doi.org/10.1016/0370-2693(90)92049-O}{\emph{Phys. Lett.} {\bf
  B234} (1990) 511}.

\bibitem{Heitger:2003nj}
{\scshape ALPHA} collaboration, J.~Heitger and R.~Sommer,
  \emph{{Nonperturbative heavy quark effective theory}},
  \href{http://dx.doi.org/10.1088/1126-6708/2004/02/022}{\emph{JHEP} {\bf 02}
  (2004) 022}, [\href{http://arxiv.org/abs/hep-lat/0310035}{{\tt
  hep-lat/0310035}}].

\bibitem{ElKhadra:1996mp}
A.~X. El-Khadra, A.~S. Kronfeld and P.~B. Mackenzie, \emph{{Massive fermions in
  lattice gauge theory}},
  \href{http://dx.doi.org/10.1103/PhysRevD.55.3933}{\emph{Phys. Rev.} {\bf D55}
  (1997) 3933--3957}, [\href{http://arxiv.org/abs/hep-lat/9604004}{{\tt
  hep-lat/9604004}}].

\bibitem{Christ:2006us}
N.~H. Christ, M.~Li and H.-W. Lin, \emph{{Relativistic Heavy Quark Effective
  Action}}, \href{http://dx.doi.org/10.1103/PhysRevD.76.074505}{\emph{Phys.
  Rev.} {\bf D76} (2007) 074505},
  [\href{http://arxiv.org/abs/hep-lat/0608006}{{\tt hep-lat/0608006}}].

\bibitem{Lin:2006ur}
H.-W. Lin and N.~Christ, \emph{{Non-perturbatively Determined Relativistic
  Heavy Quark Action}},
  \href{http://dx.doi.org/10.1103/PhysRevD.76.074506}{\emph{Phys. Rev.} {\bf
  D76} (2007) 074506}, [\href{http://arxiv.org/abs/hep-lat/0608005}{{\tt
  hep-lat/0608005}}].

\bibitem{Aoki:2001ra}
S.~Aoki, Y.~Kuramashi and S.-i. Tominaga, \emph{{Relativistic heavy quarks on
  the lattice}}, \href{http://dx.doi.org/10.1143/PTP.109.383}{\emph{Prog.
  Theor. Phys.} {\bf 109} (2003) 383--413},
  [\href{http://arxiv.org/abs/hep-lat/0107009}{{\tt hep-lat/0107009}}].

\bibitem{Follana:2006rc}
{\scshape HPQCD, UKQCD} collaboration, E.~Follana, Q.~Mason, C.~Davies,
  K.~Hornbostel, G.~P. Lepage, J.~Shigemitsu et~al., \emph{{Highly improved
  staggered quarks on the lattice, with applications to charm physics}},
  \href{http://dx.doi.org/10.1103/PhysRevD.75.054502}{\emph{Phys. Rev.} {\bf
  D75} (2007) 054502}, [\href{http://arxiv.org/abs/hep-lat/0610092}{{\tt
  hep-lat/0610092}}].

\bibitem{Blossier:2009hg}
{\scshape ETM} collaboration, B.~Blossier et~al., \emph{{A Proposal for
  B-physics on current lattices}},
  \href{http://dx.doi.org/10.1007/JHEP04(2010)049}{\emph{JHEP} {\bf 04} (2010)
  049}, [\href{http://arxiv.org/abs/0909.3187}{{\tt 0909.3187}}].

\bibitem{Sachrajda:2015rea}
C.~T. Sachrajda, \emph{{Long-distance contributions to flavour-changing
  processes}}, {\emph{PoS} {\bf LATTICE2014} (2015) 023},
  [\href{http://arxiv.org/abs/1503.01691}{{\tt 1503.01691}}].

\bibitem{Cho:2015ffa}
Y.-G. Cho, S.~Hashimoto, A.~J{\"u}ttner, T.~Kaneko, M.~Marinkovic, J.-I. Noaki
  et~al., \emph{{Improved lattice fermion action for heavy quarks}},
  \href{http://dx.doi.org/10.1007/JHEP05(2015)072}{\emph{JHEP} {\bf 05} (2015)
  072}, [\href{http://arxiv.org/abs/1504.01630}{{\tt 1504.01630}}].

\bibitem{Kaplan:1992bt}
D.~B. Kaplan, \emph{{A Method for simulating chiral fermions on the lattice}},
  \href{http://dx.doi.org/10.1016/0370-2693(92)91112-M}{\emph{Phys. Lett.} {\bf
  B288} (1992) 342--347}, [\href{http://arxiv.org/abs/hep-lat/9206013}{{\tt
  hep-lat/9206013}}].

\bibitem{Shamir:1993zy}
Y.~Shamir, \emph{{Chiral fermions from lattice boundaries}},
  \href{http://dx.doi.org/10.1016/0550-3213(93)90162-I}{\emph{Nucl. Phys.} {\bf
  B406} (1993) 90--106}, [\href{http://arxiv.org/abs/hep-lat/9303005}{{\tt
  hep-lat/9303005}}].

\bibitem{Brower:2004xi}
R.~C. Brower, H.~Neff and K.~Orginos, \emph{{M{\"o}bius fermions: Improved
  domain wall chiral fermions}},
  \href{http://dx.doi.org/10.1016/j.nuclphysbps.2004.11.180}{\emph{Nucl.Phys.Proc.Suppl.}
  {\bf 140} (2005) 686--688}, [\href{http://arxiv.org/abs/hep-lat/0409118}{{\tt
  hep-lat/0409118}}].

\bibitem{Brower:2005qw}
R.~Brower, H.~Neff and K.~Orginos, \emph{{M{\"o}bius fermions}},
  \href{http://dx.doi.org/10.1016/j.nuclphysbps.2006.01.047}{\emph{Nucl.Phys.Proc.Suppl.}
  {\bf 153} (2006) 191--198}, [\href{http://arxiv.org/abs/hep-lat/0511031}{{\tt
  hep-lat/0511031}}].

\bibitem{Brower:2012vk}
R.~C. Brower, H.~Neff and K.~Orginos, \emph{{The M\'obius Domain Wall Fermion
  Algorithm}},  \href{http://arxiv.org/abs/1206.5214}{{\tt 1206.5214}}.

\bibitem{Frezzotti:2000nk}
{\scshape Alpha} collaboration, R.~Frezzotti, P.~A. Grassi, S.~Sint and
  P.~Weisz, \emph{{Lattice QCD with a chirally twisted mass term}},
  {\emph{JHEP} {\bf 08} (2001) 058},
  [\href{http://arxiv.org/abs/hep-lat/0101001}{{\tt hep-lat/0101001}}].

\bibitem{Chen:2014hva}
{\scshape TWQCD} collaboration, W.-P. Chen, Y.-C. Chen, T.-W. Chiu, H.-Y. Chou,
  T.-S. Guu and T.-H. Hsieh, \emph{{Decay Constants of Pseudoscalar $D$-mesons
  in Lattice QCD with Domain-Wall Fermion}},
  \href{http://dx.doi.org/10.1016/j.physletb.2014.07.025}{\emph{Phys. Lett.}
  {\bf B736} (2014) 231--236}, [\href{http://arxiv.org/abs/1404.3648}{{\tt
  1404.3648}}].

\bibitem{Lin:2006vc}
H.-W. Lin, S.~Ohta, A.~Soni and N.~Yamada, \emph{{Charm as a domain wall
  fermion in quenched lattice QCD}},
  \href{http://dx.doi.org/10.1103/PhysRevD.74.114506}{\emph{Phys. Rev.} {\bf
  D74} (2006) 114506}, [\href{http://arxiv.org/abs/hep-lat/0607035}{{\tt
  hep-lat/0607035}}].

\bibitem{Creutz:1987xi}
M.~Creutz, \emph{{Overrelaxation and Monte Carlo Simulation}},
  \href{http://dx.doi.org/10.1103/PhysRevD.36.515}{\emph{Phys. Rev.} {\bf D36}
  (1987) 515}.

\bibitem{Kennedy:1985nu}
A.~D. Kennedy and B.~J. Pendleton, \emph{{Improved Heat Bath Method for Monte
  Carlo Calculations in Lattice Gauge Theories}},
  \href{http://dx.doi.org/10.1016/0370-2693(85)91632-6}{\emph{Phys. Lett.} {\bf
  B156} (1985) 393--399}.

\bibitem{Cabibbo:1982zn}
N.~Cabibbo and E.~Marinari, \emph{{A New Method for Updating SU(N) Matrices in
  Computer Simulations of Gauge Theories}},
  \href{http://dx.doi.org/10.1016/0370-2693(82)90696-7}{\emph{Phys. Lett.} {\bf
  B119} (1982) 387--390}.

\bibitem{Duane:1987de}
S.~Duane, A.~D. Kennedy, B.~J. Pendleton and D.~Roweth, \emph{{Hybrid Monte
  Carlo}}, \href{http://dx.doi.org/10.1016/0370-2693(87)91197-X}{\emph{Phys.
  Lett.} {\bf B195} (1987) 216--222}.

\bibitem{DelDebbio:2004xh}
L.~Del~Debbio, G.~M. Manca and E.~Vicari, \emph{{Critical slowing down of
  topological modes}},
  \href{http://dx.doi.org/10.1016/j.physletb.2004.05.038}{\emph{Phys. Lett.}
  {\bf B594} (2004) 315--323},
  [\href{http://arxiv.org/abs/hep-lat/0403001}{{\tt hep-lat/0403001}}].

\bibitem{Schaefer:2010hu}
{\scshape ALPHA} collaboration, S.~Schaefer, R.~Sommer and F.~Virotta,
  \emph{{Critical slowing down and error analysis in lattice QCD simulations}},
  \href{http://dx.doi.org/10.1016/j.nuclphysb.2010.11.020}{\emph{Nucl. Phys.}
  {\bf B845} (2011) 93--119}, [\href{http://arxiv.org/abs/1009.5228}{{\tt
  1009.5228}}].

\bibitem{Flynn:2015uma}
J.~Flynn, A.~J{\"u}ttner, A.~Lawson and F.~Sanfilippo, \emph{{Precision study
  of critical slowing down in lattice simulations of the $CP^{N-1}$ model}},
  \href{http://arxiv.org/abs/1504.06292}{{\tt 1504.06292}}.

\bibitem{Neuberger:1997bg}
H.~Neuberger, \emph{{Vector - like gauge theories with almost massless fermions
  on the lattice}},
  \href{http://dx.doi.org/10.1103/PhysRevD.57.5417}{\emph{Phys. Rev.} {\bf D57}
  (1998) 5417--5433}, [\href{http://arxiv.org/abs/hep-lat/9710089}{{\tt
  hep-lat/9710089}}].

\bibitem{Neuberger:1997fp}
H.~Neuberger, \emph{{Exactly massless quarks on the lattice}},
  \href{http://dx.doi.org/10.1016/S0370-2693(97)01368-3}{\emph{Phys. Lett.}
  {\bf B417} (1998) 141--144},
  [\href{http://arxiv.org/abs/hep-lat/9707022}{{\tt hep-lat/9707022}}].

\bibitem{Antonio:2008zz}
{\scshape RBC, UKQCD} collaboration, D.~J. Antonio et~al., \emph{{Localization
  and chiral symmetry in three flavor domain wall QCD}},
  \href{http://dx.doi.org/10.1103/PhysRevD.77.014509}{\emph{Phys. Rev.} {\bf
  D77} (2008) 014509}, [\href{http://arxiv.org/abs/0705.2340}{{\tt
  0705.2340}}].

\bibitem{Jansen:1992tw}
K.~Jansen and M.~Schmaltz, \emph{{Critical momenta of lattice chiral
  fermions}}, \href{http://dx.doi.org/10.1016/0370-2693(92)91335-7}{\emph{Phys.
  Lett.} {\bf B296} (1992) 374--378},
  [\href{http://arxiv.org/abs/hep-lat/9209002}{{\tt hep-lat/9209002}}].

\bibitem{Kaplan:2009yg}
D.~B. Kaplan, \emph{{Chiral Symmetry and Lattice Fermions}},  in \emph{{Modern
  perspectives in lattice QCD: Quantum field theory and high performance
  computing. Proceedings, International School, 93rd Session, Les Houches,
  France, August 3-28, 2009}}, pp.~223--272, 2009.
\newblock \href{http://arxiv.org/abs/0912.2560}{{\tt 0912.2560}}.

\bibitem{Christ:2004gc}
N.~H. Christ and G.~Liu, \emph{{Massive domain wall fermions}},
  \href{http://dx.doi.org/10.1016/S0920-5632(03)02553-2}{\emph{Nucl. Phys.
  Proc. Suppl.} {\bf 129} (2004) 272--274}.

\bibitem{Liu:2003kp}
G.-f. Liu, \emph{{Quark eigenmodes and lattice QCD}}.
\newblock PhD thesis, Columbia U., 2003.

\bibitem{Blum:2014tka}
{\scshape RBC, UKQCD} collaboration, T.~Blum et~al., \emph{{Domain wall QCD
  with physical quark masses}},  \href{http://arxiv.org/abs/1411.7017}{{\tt
  1411.7017}}.

\bibitem{Curci1983a}
G.~Curci, P.~Menotti and G.~Paffuti, \emph{{Symanzik's Improved Lagrangian for
  Lattice Gauge Theory}},
  \href{http://dx.doi.org/10.1016/0370-2693(83)91043-2}{\emph{Phys.Lett.} {\bf
  B130} (1983) 205}.

\bibitem{Luscher:1984xn}
M.~L{\"u}scher and P.~Weisz, \emph{{On-Shell Improved Lattice Gauge Theories}},
  \href{http://dx.doi.org/10.1007/BF01206178}{\emph{Commun. Math. Phys.} {\bf
  97} (1985) 59}.

\bibitem{Edwards:2004sx}
{\scshape SciDAC, LHPC, UKQCD} collaboration, R.~G. Edwards and B.~Joo,
  \emph{{The Chroma software system for lattice QCD}},
  \href{http://dx.doi.org/10.1016/j.nuclphysbps.2004.11.254}{\emph{Nucl. Phys.
  Proc. Suppl.} {\bf 140} (2005) 832},
  [\href{http://arxiv.org/abs/hep-lat/0409003}{{\tt hep-lat/0409003}}].

\bibitem{Sanfilippo_code}
F.~Sanfilippo.
\newblock https://github.com/sunpho84/nissa.

\bibitem{Luscher:2010iy}
M.~L{\"u}scher, \emph{{Properties and uses of the Wilson flow in lattice QCD}},
  \href{http://dx.doi.org/10.1007/JHEP08(2010)071,
  10.1007/JHEP03(2014)092}{\emph{JHEP} {\bf 08} (2010) 071},
  [\href{http://arxiv.org/abs/1006.4518}{{\tt 1006.4518}}].

\bibitem{Borsanyi:2012zs}
S.~Borsanyi et~al., \emph{{High-precision scale setting in lattice QCD}},
  \href{http://dx.doi.org/10.1007/JHEP09(2012)010}{\emph{JHEP} {\bf 09} (2012)
  010}, [\href{http://arxiv.org/abs/1203.4469}{{\tt 1203.4469}}].

\bibitem{Hudspith_code}
R.~Hudspith.
\newblock https://github.com/RJhudspith/GLU.

\bibitem{Wolff:2003sm}
{\scshape ALPHA} collaboration, U.~Wolff, \emph{{Monte Carlo errors with less
  errors}}, \href{http://dx.doi.org/10.1016/S0010-4655(03)00467-3,
  10.1016/j.cpc.2006.12.001}{\emph{Comput. Phys. Commun.} {\bf 156} (2004)
  143--153}, [\href{http://arxiv.org/abs/hep-lat/0306017}{{\tt
  hep-lat/0306017}}].

\bibitem{Boyle:2014hxa}
P.~A. Boyle, \emph{{Conserved currents for M{\"o}bius Domain Wall Fermions}},
  \href{http://arxiv.org/abs/1411.5728}{{\tt 1411.5728}}.

\bibitem{Gusken:1989ad}
S.~Gusken, U.~Low, K.~H. Mutter, R.~Sommer, A.~Patel and K.~Schilling,
  \emph{{Nonsinglet Axial Vector Couplings of the Baryon Octet in Lattice
  {QCD}}}, \href{http://dx.doi.org/10.1016/S0370-2693(89)80034-6}{\emph{Phys.
  Lett.} {\bf B227} (1989) 266}.

\bibitem{Alexandrou:1990dq}
C.~Alexandrou, F.~Jegerlehner, S.~Gusken, K.~Schilling and R.~Sommer, \emph{{B
  meson properties from lattice QCD}},
  \href{http://dx.doi.org/10.1016/0370-2693(91)90219-G}{\emph{Phys. Lett.} {\bf
  B256} (1991) 60--67}.

\bibitem{Allton:1993wc}
{\scshape UKQCD} collaboration, C.~R. Allton et~al., \emph{{Gauge invariant
  smearing and matrix correlators using Wilson fermions at Beta = 6.2}},
  \href{http://dx.doi.org/10.1103/PhysRevD.47.5128}{\emph{Phys. Rev.} {\bf D47}
  (1993) 5128--5137}, [\href{http://arxiv.org/abs/hep-lat/9303009}{{\tt
  hep-lat/9303009}}].

\bibitem{Foster:1998vw}
{\scshape UKQCD} collaboration, M.~Foster and C.~Michael, \emph{{Quark mass
  dependence of hadron masses from lattice QCD}},
  \href{http://dx.doi.org/10.1103/PhysRevD.59.074503}{\emph{Phys. Rev.} {\bf
  D59} (1999) 074503}, [\href{http://arxiv.org/abs/hep-lat/9810021}{{\tt
  hep-lat/9810021}}].

\bibitem{McNeile:2006bz}
{\scshape UKQCD} collaboration, C.~McNeile and C.~Michael, \emph{{Decay width
  of light quark hybrid meson from the lattice}},
  \href{http://dx.doi.org/10.1103/PhysRevD.73.074506}{\emph{Phys. Rev.} {\bf
  D73} (2006) 074506}, [\href{http://arxiv.org/abs/hep-lat/0603007}{{\tt
  hep-lat/0603007}}].

\bibitem{Boyle:2008rh}
P.~A. Boyle, A.~J{\"u}ttner, C.~Kelly and R.~D. Kenway, \emph{{Use of
  stochastic sources for the lattice determination of light quark physics}},
  \href{http://dx.doi.org/10.1088/1126-6708/2008/08/086}{\emph{JHEP} {\bf 08}
  (2008) 086}, [\href{http://arxiv.org/abs/0804.1501}{{\tt 0804.1501}}].

\bibitem{Juttner:2005ks}
A.~J{\"u}ttner and M.~Della~Morte, \emph{{Heavy quark propagators with improved
  precision using domain decomposition}}, {\emph{PoS} {\bf LAT2005} (2006)
  204}, [\href{http://arxiv.org/abs/hep-lat/0508023}{{\tt hep-lat/0508023}}].

\bibitem{Borici:1999zw}
A.~Borici, \emph{{Truncated overlap fermions}},
  \href{http://dx.doi.org/10.1016/S0920-5632(00)91802-4}{\emph{Nucl. Phys.
  Proc. Suppl.} {\bf 83} (2000) 771--773},
  [\href{http://arxiv.org/abs/hep-lat/9909057}{{\tt hep-lat/9909057}}].

\bibitem{Borici:1999da}
A.~Borici, \emph{{Truncated overlap fermions: The Link between overlap and
  domain wall fermions}},  in \emph{{Lattice fermions and structure of the
  vacuum. Proceedings, NATO Advanced Research Workshop, Dubna, Russia, October
  5-9, 1999}}, pp.~41--52, 1999.
\newblock \href{http://arxiv.org/abs/hep-lat/9912040}{{\tt hep-lat/9912040}}.

\bibitem{Kennedy:2006ax}
A.~Kennedy, \emph{{Algorithms for dynamical fermions}},
  \href{http://arxiv.org/abs/hep-lat/0607038}{{\tt hep-lat/0607038}}.

\bibitem{Hernandez:1998et}
P.~Hernandez, K.~Jansen and M.~L{\"u}scher, \emph{{Locality properties of
  Neuberger's lattice Dirac operator}},
  \href{http://dx.doi.org/10.1016/S0550-3213(99)00213-8}{\emph{Nucl. Phys.}
  {\bf B552} (1999) 363--378},
  [\href{http://arxiv.org/abs/hep-lat/9808010}{{\tt hep-lat/9808010}}].

\bibitem{Davies:2010ip}
C.~T.~H. Davies, C.~McNeile, E.~Follana, G.~P. Lepage, H.~Na and J.~Shigemitsu,
  \emph{{Update: Precision $D_s$ decay constant from full lattice QCD using
  very fine lattices}},
  \href{http://dx.doi.org/10.1103/PhysRevD.82.114504}{\emph{Phys. Rev.} {\bf
  D82} (2010) 114504}, [\href{http://arxiv.org/abs/1008.4018}{{\tt
  1008.4018}}].

\bibitem{Agashe:2014kda}
{\scshape Particle Data Group} collaboration, K.~A. Olive et~al., \emph{{Review
  of Particle Physics}},
  \href{http://dx.doi.org/10.1088/1674-1137/38/9/090001}{\emph{Chin. Phys.}
  {\bf C38} (2014) 090001}.

\bibitem{Aoki:2013ldr}
S.~Aoki et~al., \emph{{Review of lattice results concerning low-energy particle
  physics}}, \href{http://dx.doi.org/10.1140/epjc/s10052-014-2890-7}{\emph{Eur.
  Phys. J.} {\bf C74} (2014) 2890}, [\href{http://arxiv.org/abs/1310.8555}{{\tt
  1310.8555}}].

\bibitem{Boyle:2015kyy}
P.~Boyle, L.~Del~Debbio, A.~Khamseh, A.~J{\"u}ttner, F.~Sanfilippo and J.~T.
  Tsang, \emph{{Domain Wall Charm Physics with Physical Pion Masses: Decay
  constants, bag and $\xi$ parameters}},  in \emph{{Proceedings, 33rd
  International Symposium on Lattice Field Theory (Lattice 2015)}}, 2015.
\newblock \href{http://arxiv.org/abs/1511.09328}{{\tt 1511.09328}}.

\end{thebibliography}\endgroup

\end{document}